%% file: SFusion.tex
\documentclass[10pt,journal,compsoc]{IEEEtran}
\usepackage{amsmath,amssymb,amsfonts}

\usepackage{subfigure}
\usepackage{subcaption}
\usepackage{graphicx}    
\usepackage{caption}     
\usepackage{mathtools}
\usepackage{graphicx}
\usepackage{epsfig} 
\usepackage{amsthm}
\usepackage{xcolor}

  \usepackage{cite}
\ifCLASSINFOpdf
\else
\fi
\begin{document}
%
\title{SFusion: Energy and Coding Fusion for Ultra-Robust Low-SNR LoRa Networks}
%
%
%
%

\author{Weiwei Chen, Huaxuan Xiao,Jiefeng Zhang, Xianjin Xia, Shuai Wang, Xianjun Deng, Dan Zeng

\thanks{Weiwei Chen (chen.ava.0012@gmail.com) and Huaxuan Xiao are with the School of Computer Engineering and Science, Shanghai University, Shanghai, China. }
\thanks{Jiefeng Zhang and Shuai Wang are with the School of Computer Science and Engineering, Southeast University, Nanjing, China}
\thanks{Xianjin Xia is with the Department of Computer Science, The Hong Kong Polytechnic University, Hong Kong SAR, China}
\thanks{Xianjun Deng is with the School of Cyber Science and Engineering, Huazhong University of Science and Technology, Wuhan, China.}
\thanks{Dan Zeng is with the School of Communication and Information Engineering, Shanghai University, Shanghai, China.}
}

\IEEEtitleabstractindextext{%
\begin{abstract}
LoRa has become a cornerstone for city-wide IoT applications due to its long-range, low-power communication. It achieves extended transmission by spreading symbols over multiple samples, with redundancy controlled by the Spreading Factor (SF), and further error resilience provided by Forward Error Correction (FEC). However, practical limits on SF and the separation between signal-level demodulation and coding-level error correction in conventional LoRa PHY leave it vulnerable under extremely weak signals—common in city-scale deployments. To address this, we present SFusion, a software-based coding framework that jointly leverages signal-level aggregation and coding-level redundancy to enhance LoRa’s robustness. When signals fall below the decodable threshold, SFusion encodes a quasi-SF$(k+m)$ symbol using $2^m$ SF$k$ symbols to boost processing gain through energy accumulation. Once partial decoding becomes feasible with energy aggregation, an opportunistic decoding strategy directly combines IQ signals across symbols to recover errors. Extensive evaluations show that SFusion achieves up to 15dB gain over SF12 and up to 13dB improvement over state-of-the-art solutions.




\end{abstract}

}

\maketitle

\IEEEdisplaynontitleabstractindextext

%
\IEEEpeerreviewmaketitle

\input{sections/Introduction.tex}
\input{sections/motivations.tex}

\input{sections/design_overview}

\input{sections/encoder}
\input{sections/decoder}
\input{sections/evaluations}

\input{sections/related_work}
\input{sections/conclusion}

\ifCLASSOPTIONcaptionsoff
  \newpage
\fi

\bibliographystyle{IEEEtran}
\bibliography{reference}




\end{document}

%% file: sections/Introduction.tex
\IEEEraisesectionheading{\section{Introduction}\label{sec:introduction}}
LoRa \cite{devalal2018lora}, a leading Low Power Wide Area Network (LPWAN) technology, enables long-range, low-power wireless communication, making it ideal for large-scale Internet of Things (IoT) deployments. Its adoption spans diverse application domains—including agricultural monitoring \cite{ji2019lora}, wildlife tracking \cite{panicker2019lora}, and marine sensing \cite{radeta2020loraquatica}—where communication ranges often extend to tens of kilometers.

LoRa achieves long-range communication through Chirp Spread Spectrum (CSS) modulation, where each symbol is spread across $2^k$ samples (also named as chips), with $k$ denoting the Spreading Factor (SF). In general, higher SFs offer greater processing gains and resilience against noise, but at the cost of increased transmission time and energy. LoRa PHY \cite{semtech2020sx1276} specifies that SF ranges from 6 to 12. 

In addition LoRa PHY also provides Forward Error Correction (FEC) code with different Coding Rates (CR) to enhance the reception robustness. Particularly, the available CRs in LoRa PHY are CR4/5, CR4/6, CR4/7 and CR4/8, indicating that four symbols are encoded with five to eight symbols. The SF value and the CR together play pivotal roles in adapting LoRa's physical layer (LoRa PHY) to varying link conditions. 

Nevertheless, the limited SF and CR severely affects LoRa’s robustness under extremely weak link conditions. These limitations are especially critical in urban or obstructed environments, where walls, vehicles, and other obstacles significantly attenuate signal strength. This is especially critical for low-battery devices using the Class-A MAC protocol, which do not retransmit any uplink message until acknowledged by the gateway within a specified timeframe \cite{lora1ts001}, leading to extremely long delays and unwarranted battery consumption. This raises a fundamental question:
How can we design an efficient and flexible coding strategy compatible to LoRa transmitters to improve signal robustness under harsh channel conditions? Fortunately, we observe two key insights:

\textbf{1.	SF as Repetition-Like Energy Accumulation:} The SF mechanism in LoRa operates similarly to a repetition code in the time-frequency domain. Increasing SF spreads a symbol across more chips, increasing the chance of correct detection under low SNR by accumulating more energy per bit.

\textbf{2.	Symbol-Level Recovery with Opportunistic Decoding:} While repetition-based schemes improve robustness, they become inefficient when some symbols can already be decoded correctly as repetitions bring in high coding redundancy. In such cases, it is more effective to exploit the correctly decoded symbols to assist in recovering the erroneous ones, using inter-symbol coding relationships.

Motivated by the insights, we propose SFusion—a joint encoding and decoding paradigm that extends LoRa’s robustness beyond its conventional SF limit, while maintaining compatibility with commercial off-the-shelf (COTS) hardware. At the core of SFusion is a lightweight software-based encoding strategy that modulates a quasi-SF$(k+m)$ symbol (also named as SFusion symbols) using $2^m$ standard SF$k$ symbols. Here, $m$ is a tunable parameter, allowing flexible adaptation to the link quality. This approach bypasses the SF=12 limitation of the LoRa PHY by effectively constructing longer-duration symbols in software, enabling robust communication under extremely low SNR. Importantly, SFusion retains the structure required by LoRa PHY, ensuring full compatibility with its Forward Error Correction (FEC) schemes. This allows SFusion to harness built-in error correction while layering additional redundancy for more reliable decoding. 

To further enhance robustness, we introduce an opportunistic decoding strategy that processes symbols in a block as a whole, rather than demodulating them individually. By leveraging the coding correlations within the block, we can combine the energy from multiple weak SFusion symbols. This approach \textbf{transforms bit-level coding correlations into signal-level energy enhancement}, allowing erroneously demodulated symbols to be recovered by boosting their received energy through their correlations with correctly demodulated symbols.






Implementing SFusion introduces two main challenges. Firstly, how to enable efficient energy combining in the presence of hardware imperfections is nontrivial. Coherent combination of energy across the $2^m$ SF$k$ symbols that make up a quasi-SF$(k+m)$ symbol is essential for maximizing processing gain. However, hardware imperfections such as time-varying phase and frequency offsets distort the received signal and complicate combining—particularly under low SNR conditions \cite{zhang2022const}. To address this, we design and insert specially crafted \textbf{PILOT symbols} within each packet. When processed jointly with the LoRa preamble, these PILOTs enable accurate estimation and compensation of hardware-induced offsets. To minimize their overhead, we further propose an adaptive PILOT placement strategy that balances estimation accuracy with payload efficiency.

Secondly, efficient use of redundancy via grouped repetition coding requires substantial efforts. To exploit the coding structure of LoRa PHY more effectively, we develop a grouped repetition code. As illustrated in Fig.~\ref{fig:super-group}, instead of simply repeating individual symbols, we organize payload data into groups (e.g., four-symbol blocks) and repeat these groups $2^m$ times. This structure introduces redundancy at the group level and enables the novel opportunistic decoding algorithm, which uses correctly received symbols and intra-group correlations to reconstruct erroneous ones.


SFusion enables versatile transmission rates. For instance, a quasi-SF$(12+1)$ packet with CR4/6 can be emulated as an SF12 packet with CR4/6 by repeating its symbol groups twice. Notably, this is the first work to support extremely high SF for COTS devices. Furthermore, it operates independently and can be integrated with other methods that enhance LoRa communications in challenging channel conditions \cite{dongare2018charm, zhang2022const, balanuta2020cloud, mao2023recovering, mei2023ecrlora, liu2020nephalai, marcelis2017dare, chen2020exploiting, yazdani2022divide, yang2022lldpc} for further improvement. In summary, the key contributions are as follows.

1. We propose SFusion, a joint encoding and decoding paradigm, that enables signal level fusion across symbols. SFusion elegantly facilitates energy combining across: i). chirps carrying the same symbol; and ii) opportunistically boosts the energy of the incorrectly received symbols via mandatory coding constraints in LoRa PHY. Thus, SFusion significantly extends the receiver’s resilience to noise or deep fading channels.
    
2. We implement SFusion on COTS LoRa transmitter. Particularly, we leverage reverse engineering to perform pilot insertion and grouped repetition in LoRa PHY by manipulating the payload. The pilots help estimate and compensate for various signal offsets, thus ensuring coherent combining of the received signals. The receiver platform then exploits energy combing to firstly demodulate quasi-SF symbols with high confidence. Then it jointly processes these high-confidence symbols to recover other low-confidence symbols in the same coding block. 






3. We conduct extensive experiments on commodity LoRa nodes (i.e. Semtech SX1278) to evaluate SFusion, and compare it with existing studies. The results indicate that SFusion achieves up to a 13dB SNR gain compared to existing works.

The rest of the paper is organized as follows. Sec. \ref{motivation} provides the background and motivations. Sec. \ref{overview} presents SFusion's system architecture. Encoder and decoder designs are elaborated in Sec. \ref{encoder} and Sec. \ref{decoder1}, respectively. Extensive evaluations are reported in Sec. \ref{evaluation}. Related works are reviewed in Sec. \ref{relatedworks}. The paper is concluded in Sec. \ref{conclusion}.

%% file: sections/motivations.tex
\section{Background and Motivations}\label{motivation}

This section presents an overview of common challenges in LoRa communications and the motivations behind this work. For the sake of clarity and to maintain consistency, this paper uses the terms "LoRa symbol" and "LoRa chirp" interchangeably.

\subsection{A Primer on LoRa}
LoRa employs Chirp Spread Spectrum (CSS) for modulation, expanding each symbol into multiple chips, the number of which is determined by the SF. All chips within a chirp traverse the entire bandwidth, and the initial frequency of a chirp (or the frequency of the first chip within a chirp) modulates the transmitting symbol. In a LoRa packet, the $s$th chirp is denoted by $z_s$, and the $i$th chip of the $s$th chirp (represented as $x_{s,i}$) is as follows:
\begin{equation}\label{eq1}
	x_{s,i}=\begin{cases}
		e^{j2\pi (\frac{i}{2}+z_{s}-\frac{N}{2})\frac{i}{N}} & 0\leq i< N-z_{s} \\
		e^{j2\pi (\frac{i}{2}+z_{s}-\frac{3N}{2})\frac{i}{N}} & N-z_{s}\leq i< N.
	\end{cases}
\end{equation}
As the frequency of a chirp reaches its maximum value, it switches the frequency to the lowest value, resulting in a frequency discontinuity at the $(N-z_s)$th chip. We refer to chips 0 through $N-z_s-1$ as the \textbf{first chip set} of a symbol, and chips $N-z_s$ through $N-1$ as the \textbf{second chip set}. By de-chirping the signal using a standard down-chirp sequence, the chips within a chirp are transformed into a single-frequency tone. This frequency tone is subsequently extracted using an FFT operation.

\subsection{The Impact of Frequency Leakage}

In practice, there exist two types of frequency leakage that can drastically affect decoding efficiency. 

\subsubsection{Frequency leakage brought by channel variations}
 LoRa PHY employs different SF values to accommodate varying channel conditions, with a maximum SF equals 12. Consequently, links that cannot be sustained using SF12 packets may encounter communication restrictions. A straightforward approach to improve link robustness is to increase the SF arbitrarily. However, raising the SF without proper consideration can introduce practical issues, potentially reducing the expected processing gain associated with the extended SF.

Here is a contributing factor. LoRa is characterized by its low bandwidth, supporting very low transmission rates for the chips. For example, with a 125kHz transmission bandwidth, the chip duration is 8$\mu$s. The duration of the symbol for an SF12 symbol extends beyond 32ms, and if an SF16 packet is used, the duration of the symbol exceeds 524 ms. Furthermore, the typical coherence time of a wireless channel is several tens of milliseconds \cite{tse2005fundamentals}. Beyond this duration, the complex real-world environment along the lengthy transmission path, which spans kilometers, is deemed to undergo significant variations. It implies that an SF16 symbol can undergo rapid channel variations, which prevents chips within a symbol from being coherently combined, ultimately resulting in a significant reduction in the processing gain associated with a higher SF.

\begin{figure*}
    \centering       
    \subfigure[Spectrum leakage with channel variation]{
		\includegraphics[width=0.23\linewidth]{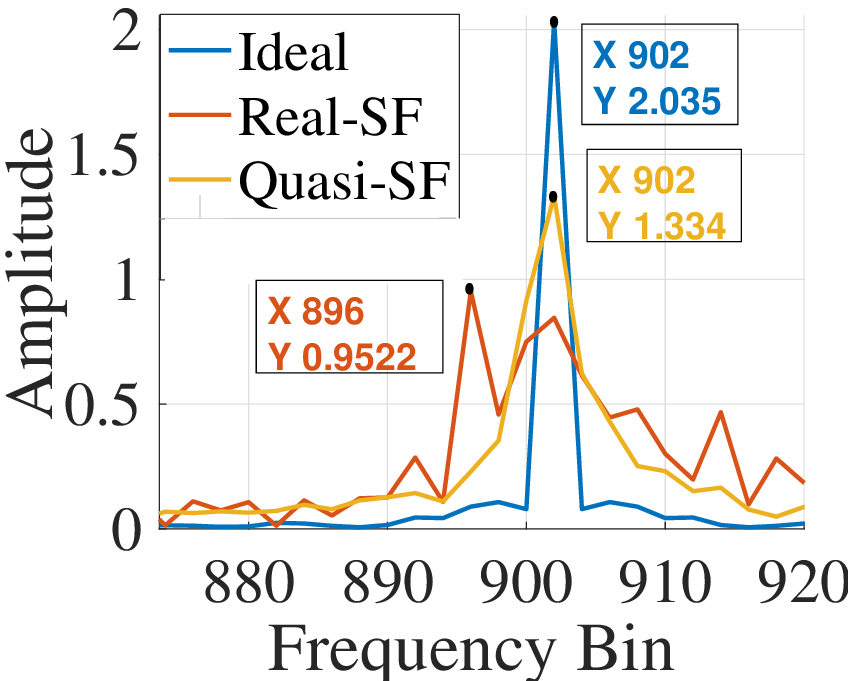}
        \label{fig:quasi-sf}    
	}
     \subfigure[Phases of each chip]{
        \includegraphics[width=0.23\linewidth]{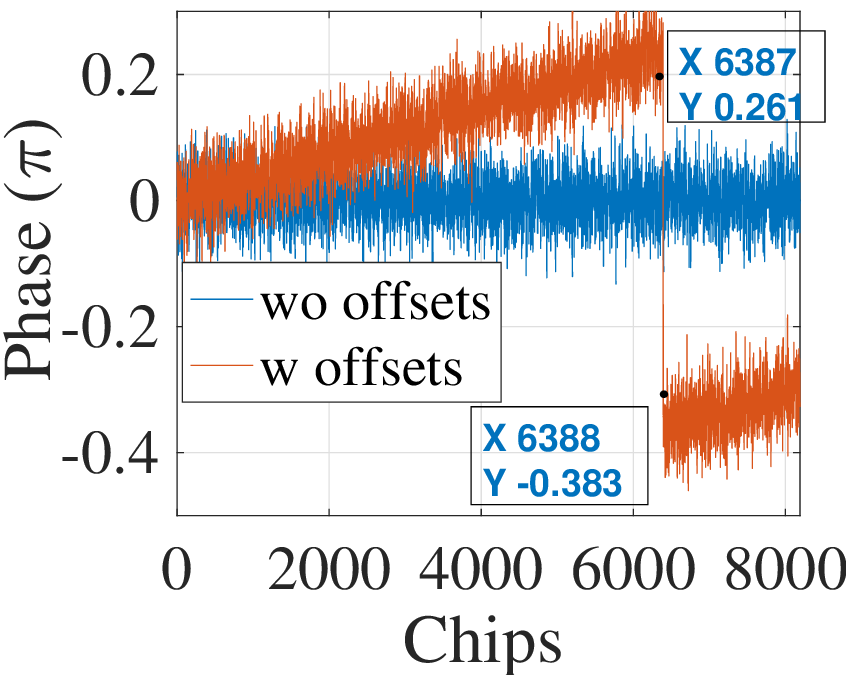}
        \label{fig:intra-off}    
    }
    \subfigure[Spectrum leakage with STO]{
    \includegraphics[width=0.23\linewidth]{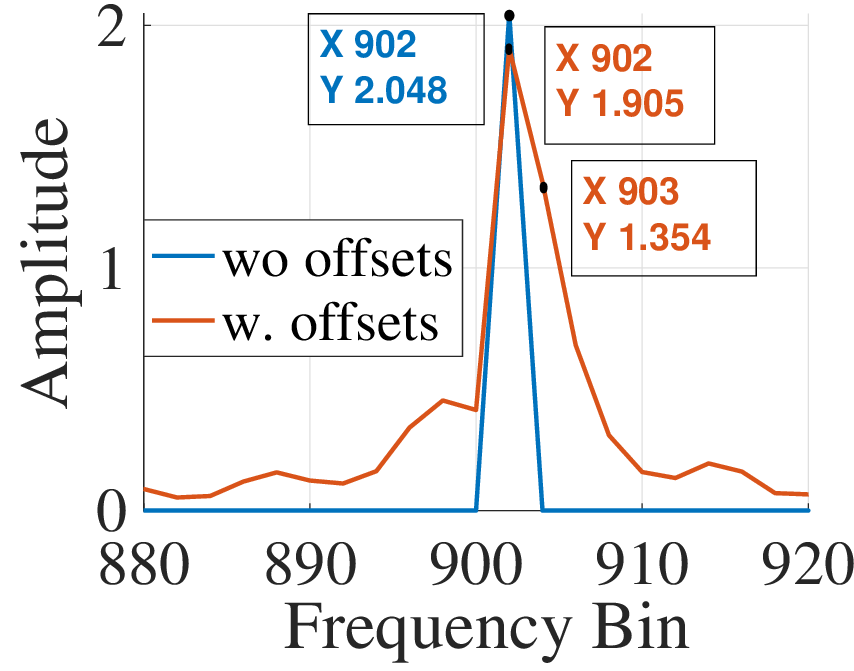}
    \label{fig:double-sample-rate}}
    \subfigure[Spectrum leakage with frequency drift]{\includegraphics[width=0.25\linewidth]{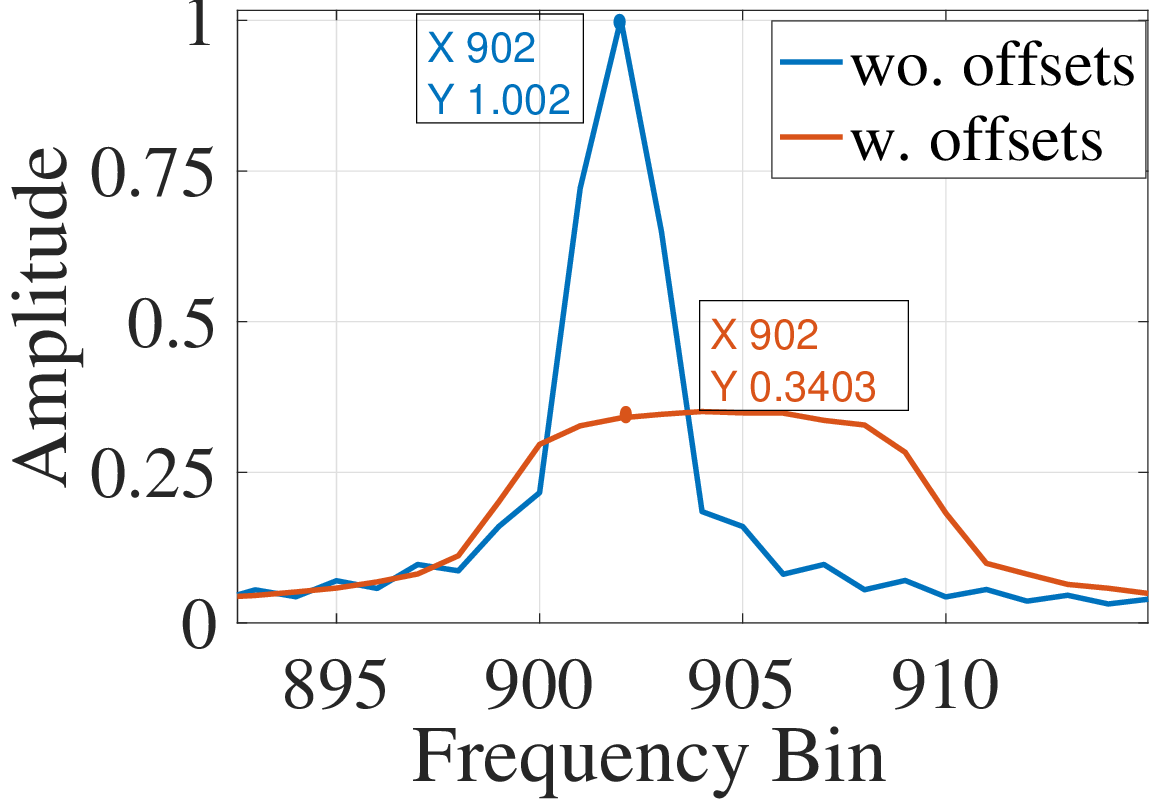}
    \label{fig:frequencyflattern}}
    \caption{(a). The spectrum distribution of a real SF14 symbol, a Quasi-SF$(12+2)$ symbol, and the ideal SF14 symbol with ideal compensation for channel state and hardware imperfections. (b). The phases of each chip in a symbol after removing the starting frequency $z_s=902$. (c) The spectrum of an SF12 symbol with and without intra-chirp offsets. (d). An illustration of how frequency drifts degrade the signal combining efficiency.}
    \label{fig:ch}
\end{figure*}

Fig. \ref{fig:quasi-sf} presents the spectrum distributions of an SF14 symbol with or without ideal combining, and that of a quasi-SF$(12+2)$ symbol. As COTS device cannot generate SF14 symbols, here both the SF14 symbol and the quasi-SF$(12+2)$ are generated and transmitted by a USRP radio. The ideally combined SF14 symbol results in a highest peak. The quasi-SF$(12+2)$ symbol generates a frequency peak located in the same frequency bin as the one generated from the ideal SF14 symbol with slightly lower amplitude. However, the SF14 symbol without compensation, suffering from frequency leakage, hardly concentrates the energy into any distinct peaks. Ideally, phase misalignment between chips due to channel variations can be effectively compensated for, leaving a distinguishable peak in frequency domain. However, it is impractical, since predicting the channel states for an extremely low SNR symbol is not feasible.


Fortunately, we observe that replacing an extremely long SF symbol with a set of shorter SF symbols can effectively mitigate the frequency leakage caused by channel variations. Take SF14 as an example, a quasi-SF$(12+2)$ symbol simulating an SF14 symbol is composed of $2^2$ SF$12$ symbols. To demodulate such a symbol, we first leverage the standard LoRa PHY approach to combine chips within each SF$12$ symbol. We aggregate the absolute values of the four symbols to obtain the final demodulation. Since all chips in the SF$12$ symbols are well-aligned (given the relatively stable channel state within an SF$12$ symbol), the frequency leakage within a symbol is greatly reduced, facilitating accurate decoding.

\subsubsection{Frequency leakage brought by hardware offsets}
As demonstrated in \cite{zhang2022const}, hardware imperfections significantly distort the received signal. Typically, these imperfections fall into three categories: Carrier Frequency Offset (CFO), Sampling Time Offset (STO), and Sampling Frequency Offset (SFO). Alongside the effects of channel variations, hardware imperfections inevitably introduce frequency leakages when receiving high SF symbols.

    \textbf{Phase offsets within a symbol}: Fig. \ref{fig:intra-off} displays the phases of all the chips within a symbol received by both an ideal receiver and an actual receiver. In this scenario, an SF12 symbol is transmitted with a bandwidth of 125kHz, and the sampling rate is twice that of the bandwidth. The transmitted symbol is $z_s=902$. Notably, a phase shift occurs between the first chip set (chip 0 to 6387) and the second chip set (the remaining chips). Fig. \ref{fig:double-sample-rate} highlights frequency leakage in FFT results due to hardware imperfections, revealing other peaks (e.g., 900, 901, 903, 904) with amplitudes close to the main peak at 902, indicating significant signal distortion. This leakage, especially in low SNR packets, can be exacerbated by environmental noise, potentially drowning out the intended signal.

    

    \textbf{Continuous frequency drift across symbols}: As shown in Fig. \ref{fig:intra-off}, the slope of the red lines represents the frequency drift within a symbol, a joint consequence of both STO and CFO as discussed in \cite{zhang2022const}. Due to the SFO, the STO experiences changes over time, resulting in a time-varying frequency drift for each symbol. Although frequency drift is negligible within one symbol, it accumulates over the course of receiving a complete packet. According to the datasheet \cite{semtech2020sx1276}, the allowable frequency drift for long SF packets (in which LDRO mode is recommended) is up to $\frac{16}{3}$ chips.
  
    Employing repetition codes further magnifies this challenge. The higher the repetition rate, the longer the packet length, and the more pronounced the frequency drifts. Therefore, repeating symbols in a packet, without compensating for offsets as is done in Ostinato \cite{xu2022ostinato}, limits the highest possible repetition rate due to ever-growing frequency drifts. Once the frequency drifts exceed one frequency bin, the subsequent symbols cannot be demodulated correctly, leading to constant bit errors regardless of SNR levels.
    
When combining SF$k$ symbols to emulate a quasi-SF$(k+m)$ symbol, it signifies that the frequency peaks of each SF$k$ symbol will not align, but will drift over time. Consequently, instead of consolidating the energy of the frequency peaks, it spreads the energy, introducing substantial frequency leakage. Fig. \ref{fig:frequencyflattern} depicts the impact of frequency drifts when combining 32 symbols. In this case, the ever-growing frequency drift not only shifts target peak (bin 902), but causes severe frequency leakage which spreads the peak energy into a wide range of frequency bins (bin 900 to 910).

Therefore, to achieve the processing gain when emulating SFusion symbols, hardware imperfections need to be meticulously compensated for. However, accurately estimating frequency and phase offsets becomes challenging given extremely low SNR levels. Additionally, as the length of the packet increases, the frequency drift and STO resulting from SFO tend to change over time, making it difficult to accurately estimate based solely on the preamble information. SFusion resolves this issue via pilot insertion to track these offsets throughout the packet.


\subsection{Energy enhancement with Coding Correlations}

  \begin{figure}
    \centering
    \includegraphics[width=1\linewidth]{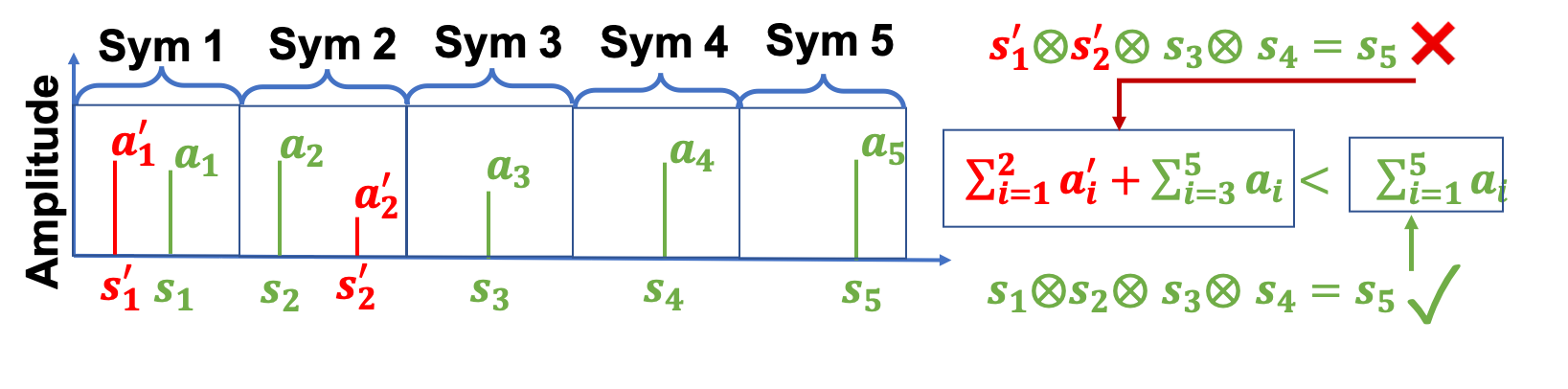}
    \caption{An illustration of joint demodulation and decoding.}
    \label{fig:joint_decode}
\end{figure}

In conventional LoRa PHY approach, the system traditionally selects the highest frequency peak in the FFT result as the demodulated symbol. Subsequently, the demodulated symbols undergo verification in the Hamming decoding module. However, independent demodulation and decoding processes underutilize the potential of the energy distribution of received IQ signals. 

Denote the set of symbols in a symbol block as $S^*$, and the amplitude of each symbol $s_i\in S^*$ be $a_{i,s_i}$. In theory, the overall amplitude of a valid symbol block should be maximized. This implies that, instead of demodulate symbols one by one, the maximum likelihood codeword should be determined as
\begin{subequations}\label{eq1}
\begin{align}
     &S^*=\arg\max_{S}\sum_{s_i\in S}a_{i,s_i}.\\
     \text{s.t.} \ \ & S \  \text{is a valid symbol block}
\end{align}    
\end{subequations}
(\ref{eq1}) clearly demonstrates that decoding symbols in a block-wide manner enables energy combining across different symbols, thus improve the overall SNR of symbols in a block. 

As an illustrative example, consider Fig. \ref{fig:joint_decode}, where CR4/5 is used and the symbol block consists of five symbols, denoted as $S=\{s_1,s_2,s_3,s_4,s_5\}$. For example, $s_1\oplus s_2\oplus s_3\oplus s_4=s_5$. Now, consider another symbol block $S'=\{s'_1,s'_2,s_3,s_4,s_5\}$ that also satisfies the coding constraints, but is incorrect. Let $a_k$ represent the amplitude of the correct symbol $s_k$, and $a'_k$  represent the amplitude of the incorrect symbol $s'_k$. Although $a'_1>a_1$ as shown in the figure, the block-wide amplitude (i.e., the total amplitude of all symbols within the block) of symbol block $S$ is larger than that of $S'$. Therefore, we should decode the symbol block as $S$, not $S'$.

%% file: sections/design_overview.tex
\section{System Overview}\label{overview}

\begin{figure}
    \centering
    \includegraphics[width=1\linewidth]{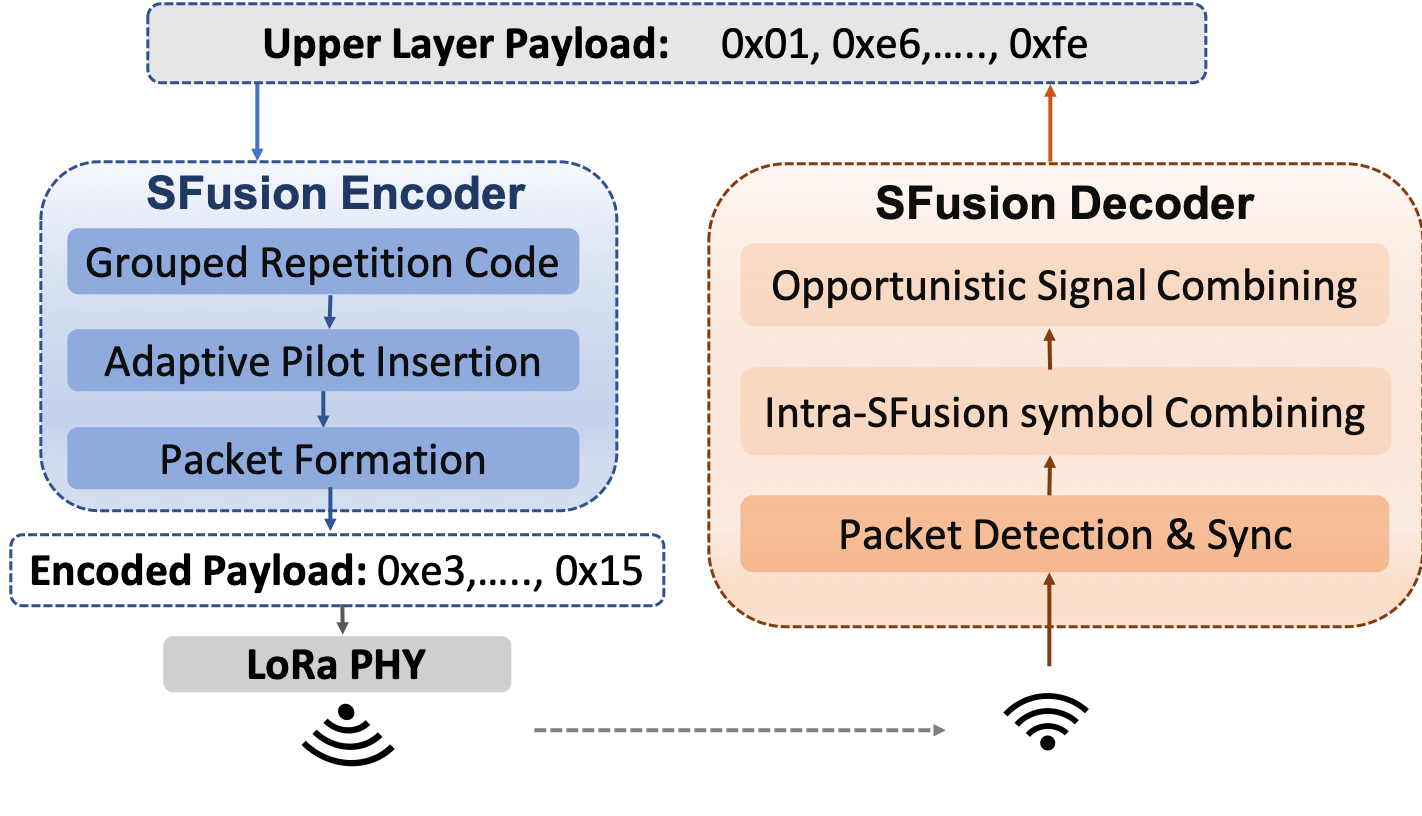}
    \caption{The overall design architecture of SFusion.}
    \label{fig:arch}
\end{figure}

As shown in Fig. \ref{fig:arch}, the architecture of SFusion consists primarily of an encoder at the transmitter and a decoder at the receiver. 

\subsection{SFusion Encoder}
The SFusion encoding module in the transmitter proposes a \textbf{Grouped Repetition Code} to encode the upper-layer payload and generate SFusion symbols while maintaining the LoRa-PHY Hamming-like coding structure. An \textbf{Adaptive Pilot Insertion} sub-module is then incorporated to facilitate accurate hardware offset estimation at the receiver. The SFusion packet is then organized in \textbf{Packet Formation} module to streamline the transmission of SFusion packets. Notably, this encoding module operates by manipulating payload bits before transmitting them via LoRa PHY, eliminating the need for any hardware modifications to implement SFusion based encoder in a LoRa transmitter. The encoder is typically utilized in battery-constrained remote LoRa nodes. Gateway nodes, which are power-sufficient, can utilize higher power levels to reach remote nodes.

\subsection{SFusion Decoder}
The SFusion decoder is consisted of three components. The \textbf{packet detection and synchronization} module serves to detect new packet arrivals and synchronize with them. It is also responsible to extract basic packet parameters. The \textbf{intra-SFusion symbol combining} module first combines the repeated $2^m$ SF$k$ symbols into quasi-SF$(k+m)$ symbols (SFusion symbols). The \textbf{Opportunistic Signal Combining} module introduces a block-wide decdoing strategy that leverages the high-confidence SFusion symbols to recover the low-confidence SFusion symbols. The output is the 0-1 payload bits.






%% file: sections/encoder.tex
\section{SFusion Encoder Design}\label{encoder}

This section describes the SFusion packet format and how to encode the upper-layer payload into an SFsuion packet.

\subsection{Grouped Repetition Code}
In LoRa PHY, four symbols form a sub-block. Symbols in a sub-block will be encoded by Hamming code to produce a larger symbol block.  For example, when coding rate is CR4/5 or CR4/7, four symbols are encoded into a block of five or seven symbols, with the first four symbols being the payload and the rest being parity symbols.

Since the Hamming coding structure in LoRa PHY is mandatory, instead of wasting the coding redundancy, as done in Ostinato \cite{xu2022ostinato}, SFusion encoder tries to fully leverage the coding structure to further improve the robustness. To achieve this, both the energy of the payload symbols and the parity symbols should be enhanced. SFusion thus employs a grouped repetition code, bundling four symbols together and repeating them in groups. This approach also leverages time diversity by distributing chips comprising an SF-$(k+m)$ symbol across different SF$k$ symbols, which are likely to experience varying channel conditions, leading to increased channel hardening with higher repetition rates, ensuring stable performance for different symbols in a packet.
    
\begin{figure}
    \centering
    \includegraphics[width=0.95\linewidth]{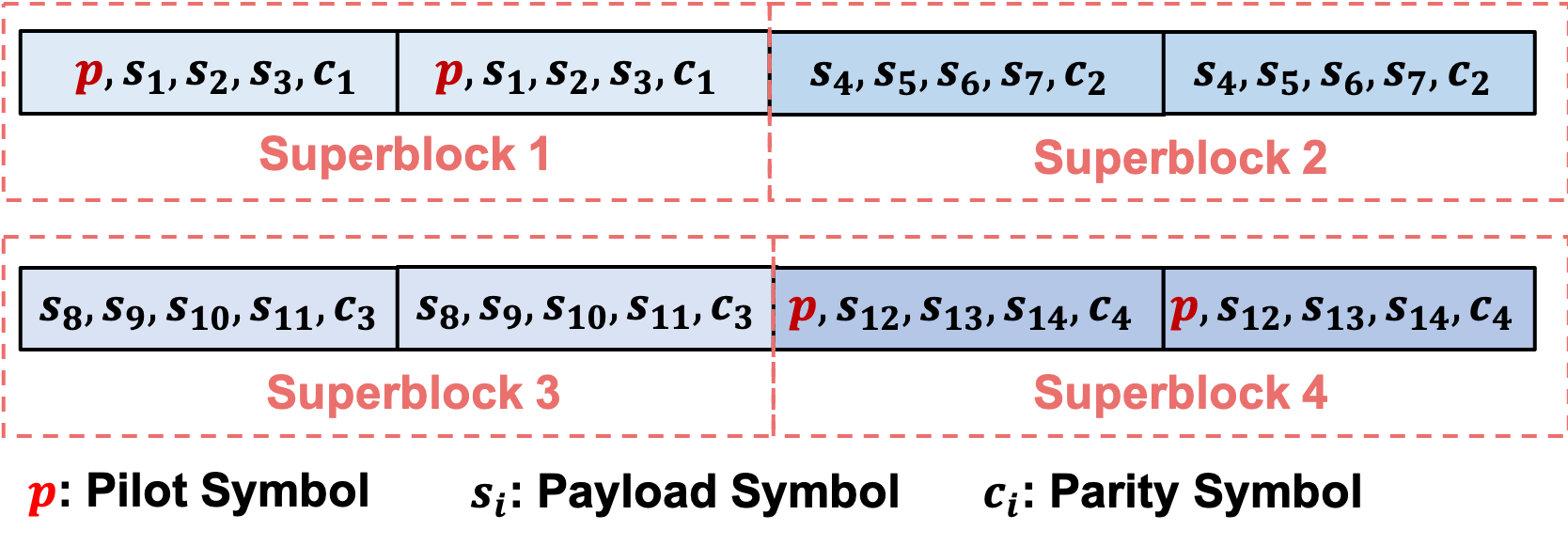}
    \caption{An illustration as how to form superblocks.}
    \label{fig:super-group}
\end{figure}

For example, when modulating quasi-SF$(k+m)$ symbols, a symbol block is repeated $2^m$ times, forming a superblock. E.g., as depicted in Fig. \ref{fig:super-group}, a packet encoded with CR4/5 and $m=1$ would comprise two blocks of symbols for a superblock.

\subsection{Adaptive Pilot Insertion}
SFusion encoder incorporates pilot symbols to enhance the tracking of hardware imperfections. These pilot symbols can take any value. In this paper, we use $z_s=2^{k-1}$ as the pilot value. As shown in the following section, setting the pilot to $z_s=2^{k-1}$ helps to achieve a balanced distribution of the number of chips in the first chip set (chip 0 to chip $N-z_s-1$) and the second chip set (chip $N-z_s$ to chip $N-1$) for the STO estimation.

\label{pilotinsertion}
A key question is how to insert pilot symbols into the packet structure. In our design, pilot symbols can be embedded in any superblock. Once inserted, each pilot is repeated $2^m$ times in accordance with the grouped repetition scheme to preserve the underlying Hamming-like coding structure.

However, since pilot symbols introduce additional transmission overhead, we propose an adaptive pilot insertion strategy to balance the tradeoff between estimation accuracy and coding efficiency. Intuitively, higher quasi-SF values lead to longer packets, which are more susceptible to phase and frequency offsets, thus requiring denser pilot placement for accurate offsets estimation.


As detailed in our evaluation (Sec.\ref{evaluation}), we empirically analyze the relationship between the number of pilots and the estimation error of hardware-induced offsets. Based on this, we derive a set of pilot placement patterns that meet the accuracy requirements while minimizing pilot overhead for each quasi-SF value.


\subsection{Packet Formation}
\begin{figure*}
    \centering
    \includegraphics[width=0.82\linewidth]{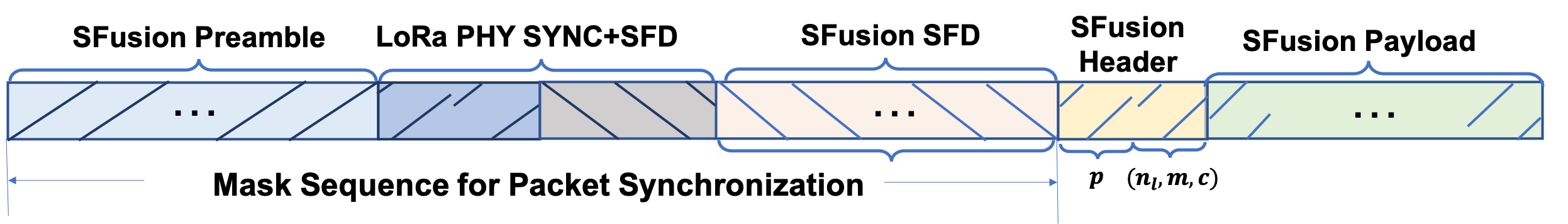}
    \caption{The structure of an SFusion packet. Here $n_l$ indicates the number of payload bytes in the SFusion packet.}     
    \label{fig:sf-format}  
\end{figure*}

SFusion is implemented on Commercial Off-The-Shelf (COTS) LoRa transmitters, and it adheres to the packet format specifications of LoRa PHY. Fig. \ref{fig:sf-format} illustrates SFusion’s packet format. Specifically, the number of preambles in a packet can be customized to an arbitrary value. When implementing a quasi-SF$(k+m)$ packet with SF$k$ symbols, the number of preambles is set to $\max\left\{8,2^{m+1}\right\}$ SF$k$ symbols to ensure accurate preamble detection. 

Furthermore, we use the implicit header mode in Semtech SX127X chipsets. In this mode, the first four symbols will be encoded with CR4/8 as done in the explicit header mode (besides the first eight symbols, the rest part of the payload will be encoded with the specified coding rate). Moreover, all the eight symbols will be transmitted with LDRO mode. This implies that, for SF$k$ packet, only $4\cdots (k-2)$ bits are encoded in the first symbol block. We insert dummy bits to the first eight symbols to generate the desired downchirp symbols for synchronizing the SFusion packet. The generation of the downchirp  symbol is similar to that of Ostinatio \cite{xu2022ostinato}. E.g., calling interruption in Frequency Hopping (FH) mode to flip the upchirps to down chirps. 



After the SFusion-based SFD, the first symbol is the pilot symbol, and the second symbol carries the (payload length $n_l$, quasi-spreading factor $m$ and coding rate $c$) information. Here we utilize the first six bits in the SFusion payload to indicate the payload length, the next three bits for quasi-spreading factor $m$, and two more bits to indicate the coding rate. The remaining bits are set to 0. In general, since SFusion packets last long, and are vulnerable to error, the SFusion packet length is restricted to 63 $(2^6-1)$ Bytes, which can be encoded with eight bits. SFusion payload will then be arranged in subsequent symbol blocks. Denote $k$ bits of SFusion header by $\left\{b_0, b_1, \cdots, b_{k-1}\right\}$, where $n_l$, $m$ and $c$ are calculated as
$n_l=\sum_{0}^{i=5}b_i\cdot 2^i$, $m=\sum_{i=6}^{8}b_i\cdot 2^{i-6}$, and $c=b_{9}+b_{10}\cdot 2$ and $b_i=0$ for any $i>10$.

Upon receiving the upper layer payload, the encoder computes the number of superblock $s_g$, and allocates pilots to their respective superblock. SFusion subsequently employs a reverse engineering technique \cite{xu2022ostinato,li2017webee} to generate the downchirp symbols and the final payload, which is then transmitted through LoRa PHY.

%% file: sections/decoder.tex
\section{SFusion Decoder Design}\label{decoder1}
This section describes how the three modules function in the decoder.

\subsection{Packet Detection and Synchronization}\label{sec:PDR}
The preamble detection mechanism identifies the arrival of a packet with sliding windows. Since the number of preambles in a quasi-SF$(k+m)$ packet is set to $\max\left\{8,2^{m+1}\right\}$, the sliding window is configured with a size of 64 ($m=5$) symbols. This window continuously scans the received signal, with a step size equal to one fourth the length of an SF$k$ symbol. In this process, all the chips within each symbol are correlated with downchirps and subjected to a standard $2^k$-point FFT. After that, four sub-windows with 8, 16, 32 and 64 SF$k$ symbols are derived to accumulate the absolute values of the spectrum for all symbols in the sub-windows, respectively. As the receiver constantly monitors environmental noise, it can detect the arrival of a new packet when the cumulative SNR of the same frequency peak from two SF$(k+m)$ chirps surpasses a predefined margin (denoted as $\Gamma_1$) in at least one sub-windows. The choice of $\Gamma_1$ is extended in the evaluations. 



Once a sliding window detects the SF$(k+m)$ preamble chirps, it further leverages a sequence of mask symbols to determine whether the newly detected symbol is a start of the packet and synchronizes to the packet if yes. This is performed as follows. The mask sequence, as indicated in Fig. \ref{fig:sf-format},  consists of the $\max\left\{8,2^{m+1}\right\}$ SFusion preamble symbols, two LoRa-PHY SYNC words, 2.25 LoRa-PHY SFD and eight downchirp symbols after the SFD (SFusion SFD). It then calculates the autocorrelation of the conjugate of the mask sequence and the received signal to finalize the arrival of a new packet, and the starting symbol of the SFusion payload. 

\subsection{SFusion packet parameter Extraction}
To extract the packet parameter, the initial step involves locating the positions of the pilot symbols, which is determined by coding rates. For example, when using CR4/5 or CR4/8 encoding, the pilot symbol repeats every five or eight symbols, respectively. Therefore, various pilot intervals are tested to identify the one that maximizes the cumulative pilot energy. More specifically, let $F_{s,p}$ represent the $p$th frequency tone of the $s$th symbol, with $p$ being the intended pilot symbol and $s$ as the starting position of the pilot symbol. The search is conducted to determine the repetition parameter $m$ and the coding rate parameter $c$ (e.g., when $c=5$, the coding rate is CR4/5) subject to the following.
\begin{equation}    \label{eq_est1}
\left\{m,c\right\}=\arg\max_{\left\{m,c\right\}}\frac{1}{m}\sum_{k=0}^{2^{m}-1} |F_{s+ck,p}|.
\end{equation}

Theoretically, $m$ and $c$ will be determined from (\ref{eq_est1}). To improve the accuracy of the spreading factor estimation we proceed to decode the SFusion header, which explicitly encodes the value of $m$ and $c$. Let $F_{s,h}$ represent the $h$th frequency tone of the $s$th symbol, where $h$ encodes the SFusion parameter ($m$, $c$ and the payload length $n_l$). We define $H_{m,c}$ as the set of frequency bins that are decoded with $m$ and $c$ (e.g., $m=\sum_{i=0}^{5}b_i\cdot 2^{i-6}$ and $c=b_9+b_{10}\cdot 2$, where $b_i$ is the $i$-th bit in symbol $h$). The target quasi-SF parameter $m^*$, the CR parameter $c^*$ and the payload length $n_l^*$ is determined by:
\begin{equation}
\left\{m^*,c^*,n_l^*\right\}=\arg\max_{h\in H_m} \frac{1}{2^m}\sum_{k=0}^{2^{m}-1} |F_{s+ck,h}|.
\end{equation}

Upon determining the optimal values of $\left\{m^*,c^*,n_l^*\right\}$, the decoder starts decoding the packet in the subsequent sections.

\subsection{Intra-SFusion Symbol Combining}
This sub-section deals with signal combining of $2^m$ SF$k$ symbols to a generate SF-$(k+m)$ symbol. Prior to combining symbols, it first details how hardware imperfections affect received signals and introduces the corresponding schemes to estimate hardware offsets. Once the offsets are well compensated for, signal combining will be effectively performed subsequently.

\subsubsection{The impact of hardware imperfections on the received symbols}
As discussed in \cite{zhang2022const}, hardware imperfections can introduce significant distortions in the received signal. Of particular concern are three key offsets: CFO (Carrier Frequency Offset), STO (Sampling Time Offset), and SFO (Sampling Frequency Offset). CFO and SFO are denoted by $\delta$ and $\eta$, respectively. While $\delta$ and $\eta$ may vary gradually across different transmissions, they are relatively stable within a single packet, assuming the oscillators in both the transmitter and receiver function normally. Additionally, let $\tau_s$ represents the STO of the $s$th symbol. Due to SFO, $\tau_s$ can vary over time. Designate $\Phi_{s}$ as the phase offset resulting from hardware imperfections. 

Since an SF$k$ symbol duration is shorter than the coherent time of the channel, we use a fixed channel state $h_s$ to approximate the channel state of each chip in a chirp. It is worth noting that in cases where the channel exhibits extremely rapid variations, it is possible to reduce the value of $k$ and the symbol duration to mitigate the extent of channel variations within a symbol. Furthermore, given that LoRa nodes are typically deployed in relatively stable locations, the frequency drift resulting from channel fading is considerably lower than that caused by hardware imperfections.

Given the above analysis, the received and de-chirped signals (up-chirps signal) are denoted with $y_{s,i}$, and characterized as \cite{zhang2022const}.\begin{equation}\label{dif_phi}
y_{s,i}=\begin{cases}
    |h_s|e^{j2\pi \Phi_{s}}e^{j2\pi (z_{s}+f_{s})\frac{i}{N}}, &  0\leq i< N-z_{s}, \\
    |h_s|e^{j2\pi \hat{\Phi}_{s}}e^{j2\pi (z_{s}+f_{s})\frac{i}{N}},  & N-z_{s}\leq i< N.
\end{cases}
\end{equation}
Here, $N=2^k$ is the number of chips in an SF$k$ symbol, $|h_s|$ and $\angle h_s$ represent the amplitude and the angle of $h_s$. With this, we have
\begin{subequations}\label{dif_phi1}
\begin{align}\label{dif_off}
        &\Phi_{s}= \angle h_s+\Phi_{s,CFO}+\Phi_{s,STO},\\
        &\Phi_{s,CFO}= (s-1)\delta,\\
        &\Phi_{s,STO}=-\frac{z_s\tau_s}{N}+\frac{\tau_s}{2}+\frac{{\tau_s}^2}{2N}, \\
        &\hat{\Phi}_{s}=\Phi_{s}+\tau_s,\label{phase_o} \\
        &f_s= \delta-\tau_{s}.
\end{align}
 \end{subequations}
 
 According to (\ref{dif_off}), there are three types of drifts caused by hardware imperfections and channel state:

    \textbf{i) Inter-symbol phase offsets}: the phase offset $\Phi_{s}$ is a joint effect of the channel state $h_s$, the CFO and the STO;
    
   \textbf{ii) Intra-symbol phase offsets}: given the assumption that channel state remains stable in a symbol, the phase offsets between the first and the second chip set within a symbol is solely caused by STO (refer to (\ref{phase_o}), $\tau_s=\hat{\Phi}_{s}-{\Phi}_{s}$); 
    
    \textbf{iii) Inter-symbol frequency drifts}: while CFO causes a constant frequency drift ($\delta$), STO causes a time-varying frequency drift ($-\tau_s$). The two factors make the frequency tones ($\delta-\tau_s$) drift further and further away from the actual transmitted frequency tone. 

\subsubsection{Hardware imperfection estimation for weak links}
When finishing decoding the SFusion header, the positions of all the PILOT symbols in the packets are determined. The decoder leverages both preambles and pilot symbols for hardware imperfection estimation. Specifically, refer to (\ref{phase_o}), the STO $\tau_s$ is calculated as:
\begin{subequations}
\begin{align}
    \tau_s&=\frac{1}{2\pi}\angle (\frac{g_{s,2}}{g_{s,1}}),\\
    g_{s,1}&=\sum_{i=0}^{N-z_s-1} y_{s,i} e^{-j2\pi (z_s+\delta-\tau_s)\frac{i}{N}},\label{sb1}\\
    g_{s,2}&=\sum_{i=N-z_s}^{N-1} y_{s,i} e^{-j2\pi (z_s+\delta-\tau_s)\frac{i}{N}}.\label{sb2}
\end{align}
\end{subequations}

When $z_s=\frac{N}{2}=2^{k-1}$, the number of chips in $g_{s,1}$ and $g_{s,2}$ are balanced well. This explains why we choose $\frac{N}{2}$ as the pilot. 

Consider the set of preambles and pilots denoted by $S_p$. For each preamble, following a similar approach as used in Magnifier \cite{chen2023magnifier}, we shift the window to the middle of two preambles. This transformation effectively converts the preambles into pilot signals carried by the frequency tone $z_s=\frac{N}{2}$. To estimate the hardware imperfections, we employ the optimization approach outlined below.\begin{subequations} \label{opt_h}
   \begin{align}
   &\max_{\tau_0, \delta,\eta}\sum_{s\in S_p}|g_{s,1}+g_{s,2}e^{-j2\pi \tau_s}|\\
    \text{s.t. \ \ \ }  
               &
               \begin{cases}
               \tau_s=\tau_0+(s-1)\eta,\\ \label{sto-linear}
                g_{s,1}=\sum\limits_{i=0}^{N-z_s-1} y_{s,i} e^{-j2\pi (z_s+\delta-\tau_s)\frac{i}{N}},\\
                g_{s,2}=\sum\limits_{i=N-z_s}^{N-1} y_{s,i} e^{-j2\pi (z_s+\delta-\tau_s)\frac{i}{N}}.
               \end{cases}
    \end{align}
\end{subequations}

According to \cite{zhang2022const}, the STO of a normally functioned transmitter grows almost linearly with time. Therefore, the STO is approximated with a linear function of SFO (\ref{sto-linear}).  

\subsubsection{Hardware imperfection compensation and SFusion symbol combining}
In theory, the maximum processing gain is achieved when all chips within an SFusion symbol are coherently combined. However, practical challenges arise as the acquisition of channel state information $h_{s,i}$ is not achievable. Consequently, we do not apply compensation for the phase offsets between symbols. 

Moreover, the frequency drift gradually increases over time due to SFO, leading to substantial frequency leakage if left unaddressed. Therefore, besides the intra-symbol phase offset, inter-symbol frequency drifts should also be compensated for as follows.

\textbf{Signal combining within a chirp}: As a symbol only lasts for several mini-seconds, the channel state in a symbol is relatively stable, chips within a symbol can be aligned to concentrate the energy of the target frequency tone while mitigating the frequency leakage arising from hardware imperfections.

Specifically, we will address the intra-symbol phase offsets caused by STO, and the frequency drift caused by both CFO and STO. Once the offset information has been estimated from (\ref{opt_h}), denote the estimated STO, CFO, and SFO by $\tau_s^*, \delta^*,\eta^*$, respectively, and the corresponding aggregation of first chip set and second chip set by $g_{s,1}^*$ and $g_{s,2}^*$ respectively. Let $y_{s_z,i} = y_{s,i}$, $Z_{s_z} = z$, and $\tau_{s_z}^* = \tau_s^*$ (refer to (\ref{sb1}) and (\ref{sb2})), the signal of the $z$th frequency tone denoted by $Y_{s,z}$ can be compensated with:
\begin{equation}
Y_{s,z} = g_{s_z,1}^* + g_{s_z,2}^* e^{-j2\pi \tau_s^*}.
\end{equation}
    
The compensation helps align chips in a symbol. After compensating for the intra-symbol phase offsets, and the inter-symbol frequency drift, frequency leakage can be significantly alleviated.
    
     \textbf{Signal combining between different chirps}: As the channel state changes from symbol to symbol, and hard to estimate due to links' extremely low SNR, compensating for the phase offsets resulted from channel variations is not viable. We therefore only compensate for the frequency drifts. Specifically, different SF$k$ symbols belonging to the same quasi-SF$(k+m)$ symbol are combined by summing their absolute values. Denote the cumulative amplitude of a frequency tone $z$ of the same quasi-SF symbol by $\tilde{Y}_z$, given coding rate $c$, $\tilde{Y}_z$ is expressed as: 
     \begin{equation}
         \tilde{Y}_z =\sum_{k=0}^{2^m-1}|Y_{s+ck,z}|.
     \end{equation}
     
This helps to average out the environmental noise while enhancing the target frequency peak.

\subsection{Opportunistic Signal Combining across SFusion symbols in a symbol block}
While SFusion based per-symbol demodulation provides significant processing gain enhancement, and thus increases the symbol reception probability tremendously, the opportunistic signal combining module further leverages coding correlations to recover erroneously demodulated SFusion symbols from the correct ones. 

\begin{figure}
    \centering
    \includegraphics[width=0.9\linewidth]{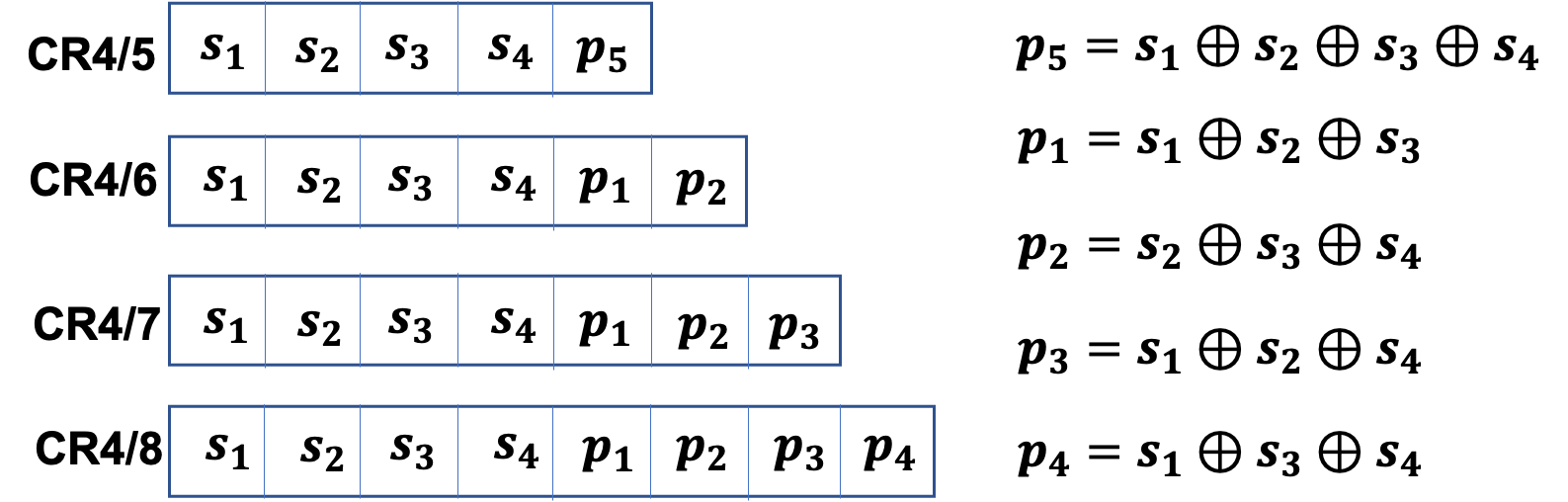}
    \caption{Hamming code used in LoRa-PHY}
    \label{fig:ham}
\end{figure}

To ease the presentation, in this subsection, a symbol refers to a SFusion symbol (combined with $2^m$ SF$k$ symbols). Recall that, the grouped repetition code maintains the Hamming-like coding structure of all the SFusion symbols. Denote $\{s_1, s_2, s_3, s_4\}$ as the set of payload symbols, $s_5$ to $s_8$ as the parity symbols based on the Hamming code employed in LoRa PHY. The generation of the parity symbols are shown in Fig. \ref{fig:ham}.

Recall from (\ref{eq1}), if a block of symbols is demodulated as a whole, all the symbol bins will be aligned jointly to maximize the accumulated amplitude of all the SFusion symbols in a superblock, e.g. $S^*=\arg\max_{S}\sum_{s_i\in S}a_{i,s_i}$, where $S$ is a valid symbol block.

Since each symbol block contains four independent symbols, the naïve solution is to exhaustively search all symbol combinations, generate the symbol block associated with each combination, and decode the symbol block as the one with the highest accumulated amplitude. However, this will incur unbearable complexity overhead. For instance, when a packet is modulated with SF=12, there will be $2^{12}$ different possible values for one symbol, and the total number of symbol combinations of a block will be $2^{48}$. 

To tackle this issue, we leverage the observation that as long as quasi-spreading factor has already improve the received signal's SNR significantly, the majority of symbol frequency peaks are higher than the noise level. Therefore, we can leverage the correct symbols to recover the error ones. In cases where most symbols cannot be decoded successfully, increasing the quasi-spreading factor is necessary to generate symbols that are more resilient to noise.

Now, given a set of high confidence received symbols (denoted $H_1$), we can find another set of symbols which are linearly independent of $H_1$ (and denoted as $H_2$). In other words, the symbols in $H_2$ cannot be represented as a linear combinations of symbols in $H_1$.
We leverage $H_1$ and the coding correlations to predict the symbols in $H_2$. Specifically, recall (\ref{eq1}) and Fig. \ref{fig:ham}, if $H_1$ is known, 
any symbol $s_i$ in the symbol block can be expressed as a function of both $H_1$ and $H_2$ with $s_i=f_i(H_1,H_2)$. As an example, suppose $H_1=\left\{s_1,s_2\right\}$, and $H_2=\left\{s_3,s_4\right\}$, then symbol $s_5$ can be expressed as $s_5=s_1\oplus s_2\oplus s_3=f_5(H_1,H_2)$. Therefore, $s_5$ is a function of $H_1$ and $H_2$. Similarly, the overall amplitude of all the symbols in a block can be expressed as
\begin{align}\label{eq_fec}
     S^*(H_1)=\arg\max_{H_2}\sum_{i=1}^{c}|a_{i,f_i(H_1,H_2)}|.
\end{align}

In essence, given $H_1$, to find the optimal symbol block $S^*(H_1)$, we need to try $2^{SF\cdot|H_2|}$ different symbol combinations. As there are four linearly independent symbols in a symbol block, $|H_1|+|H_2|=4$. Therefore, the number of combinations becomes $2^{SF\cdot(4-|H_1|)}$. For instance, if we can find four independent symbols for $H_1$, $H_2$ is empty, and the optimal symbol block can be uniquely determined by $H_1$. However, if $|H_1|=0$, $|H_4|=4$, the problem is reduced to an exhaustive search problem, in which we need to try $2^{4SF}$ different combinations. To strike the balance between complexity and decoding efficiency, we search a symbol set $H_1$ with size equal to three. In this case, $H_2$ contains one symbol, and the number of combinations is $2^{SF}$, which is equivalent to evaluating all the frequency bins in $H_2$. To summarize, the algorithm works as follows
\begin{enumerate}
    \item Select the three SFusion symbols in a coding block  with the highest accumulated values, and set them as $H_1^*$. Find a symbol that is linearly independent to $H_1^*$ and set it as $H_2$;
    \item Assign values from $[0, 2^{SF}-1]$ to $H_2$, for each value, calculate the remaining symbols in a coding block with $H_1$ and $H_2$,  and get the accumulated amplitude as $S(H_1^*,H_2)= \sum_{i=1}^{c}|a_{f_i(H_1^*,H_2)}|$;
    \item Choose the one with the highest accumulated amplitude, and set the corresponding value to the symbol in $H_2^*$.
\end{enumerate}

The above algorithm opportunistically combines signal across SFusion symbols, and effectively boosts the overall amplitude of a block of symbols.






%% file: sections/evaluations.tex
\section{Evaluations}\label{evaluation}
This sections first describes the experimental setting, and then presents the performance of SFusion.
 
\begin{figure}
\centering
    \includegraphics[width=0.8\linewidth]{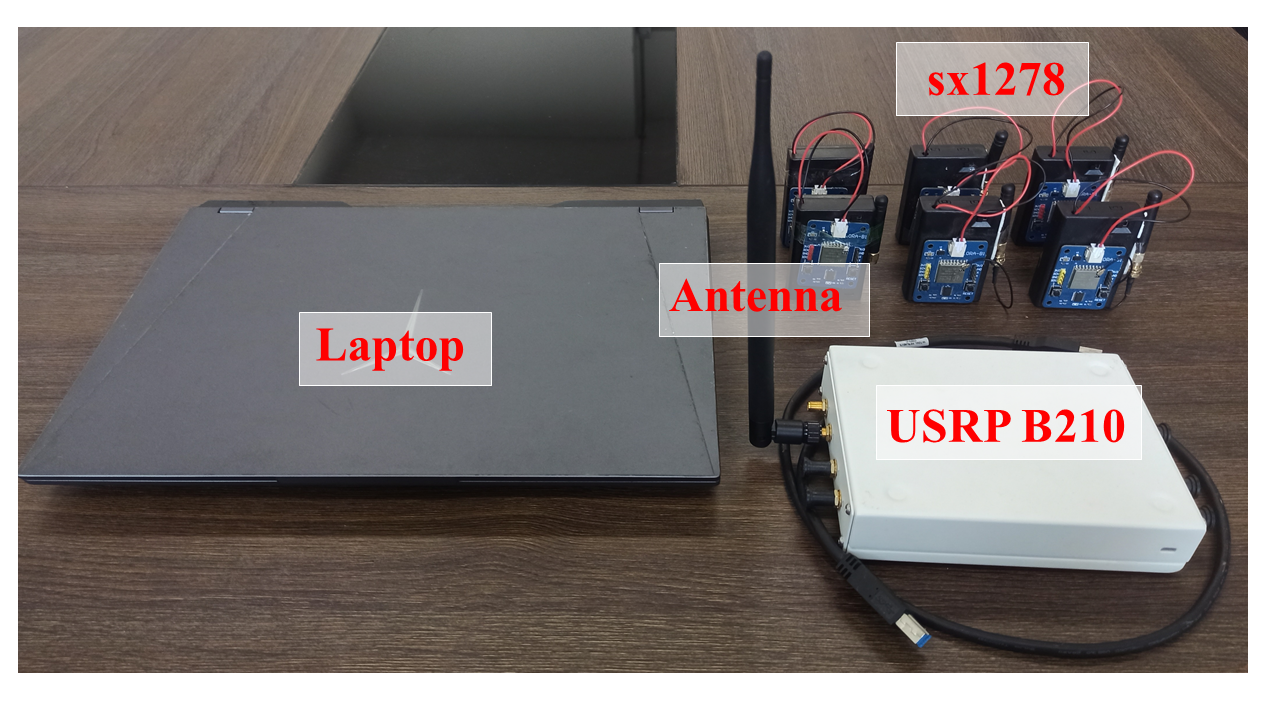}
    \vspace{-0.1in}
    \caption{ Devices used in experiments.}
    \label{fig:devices}
     \vspace{-0.2in}
\end{figure}
\subsection{Experimental Setup}\label{Setup}

     \subsubsection{The transmitter and gateway implementation.} We implement SFusion on commercial off-the-shelf (COTS) LoRa transmitters equipped with the SX1278 LoRa chip. The transmitter is controlled by MCU STM32F030F4 and powered by three 1.5V dry batteries. The communication frequency is set at 433MHz, with a bandwidth of 250kHz. The payload size for SF$(12+1)$ to SF$(12+3)$ is set to 30 bytes. The payload size for SF(12 + 4) and SF(12 + 5) packets, due to the 255 bytes limit, is set to 12 bytes and 6 bytes, respectively. 
     
     To collect the IQ signals, we use a USRP B210 as a gateway. It is connected to a laptop with an R7-5800H CPU and 32GB of RAM. The received signals are processed in Matlab for further analysis and evaluation. Note that, according to LoRa PHY specification, the maximum supportable payload for different SF schemes is different. 
     

     \subsubsection{Schemes to be compared with.} We conduct comparative analysis involving SFusion, the standard SF12 provided by LoRa PHY, Ostinato \cite{xu2022ostinato} that introduces the concept of accumulating multiple symbols for SNR enhancement, and CONST \cite{zhang2022const} that compensates for offsets and accumulates signals from multiple gateways.

     In Ostinato, a parameter denoted as $K$ is used to represent the repetition rate. Strictly speaking, the highest repetition rate Ostinato can support is 4. When $K=8$, due to coding constraints, only symbol 0 can be transmitted. In addition, Ostinato is implemented here with the Low Data Rate Optimization (LDRO) mode, as it lacks the capability to track frequency drifts within a packet. Note that, in the LDRO mode, the last two bits are disabled to mitigate the effects of frequency drifts. 
     

    Regarding CONST, the demodulation procedure consists of three parts. First, the peak energy of each symbol in different gateways is combined coarsely to select a candidate symbol list that belongs to the packet. Second, this symbol list, similar to pilot symbols in SFusion but with higher uncertainty, is used to estimate and compensate for signal offsets in the packet received by each gateway. Finally, the compensated symbols are recombined to determine the final results. To make a fair comparison, the number of gateways involved is equal to the repetition rate in Ostinato and SFusion. Similarly, the positions of the gateways are carefully selected so that the SNR of the packets received from different gateways is close.
     
     In addition, to demonstrate SFusion's compatibility with orthogonal methods, we integrate SFusion with CONST and evaluate the performance of this combined method. Here, the offset estimation and compensation parts of CONST are replaced by SFusion.
     
     \subsubsection{Evaluation Scenarios.} We evaluate both indoor and outdoor performance for various schemes. The indoor experiments explore the supportable SNR limits of different schemes. The outdoor experiments evaluate SFusion under various channel conditions. 

     \subsubsection{Performance Metrics.} We evaluate the average Bit Error Rate (BER), the Packet Reception Rate (PRR) and the effective throughput for different schemes under various situations. 
     
     
\subsection{Evaluation Results}
We evaluate SFusion with different SF and various coding rates. The outdoor experiments are also conducted to demonstrate the overall performance of SFusion. 



\subsubsection{Packet detection performance}



We begin by discussing how to determine the SNR margin (denoted as $\Gamma_1$ in Sec.~\ref{sec:PDR}) for detecting an SFusion packet. This value is critical because a high SNR margin leads to a higher rate of false negative (a weak packet cannot be detected), which degrades detection performance. On the other hand, a low margin may result in many true negatives, where noise is falsely detected as a packet, thus wasting unnecessary computational resources. Therefore, the SNR margin directly impacts overall packet detection performance.

Since the SNR margin depends on the lowest detectable SNR for a packet, we first plot the average packet detection rate, defined as the ratio of successfully detected packets to all transmitted packets. From the plot, we observe that the SNR thresholds for detecting SF$(12+1 \text{\&}2)$, SF$(12+3)$, SF$(12+4)$, and SF$(12+5)$ packets are -31 dB, -33 dB, -34 dB, and -36 dB, respectively. As shown in the following subsections (see Fig.~~\ref{fig:SNRimprovement}).



\begin{figure}
    \centering    
    \subfigure[]{
		\includegraphics[width=0.47\linewidth]{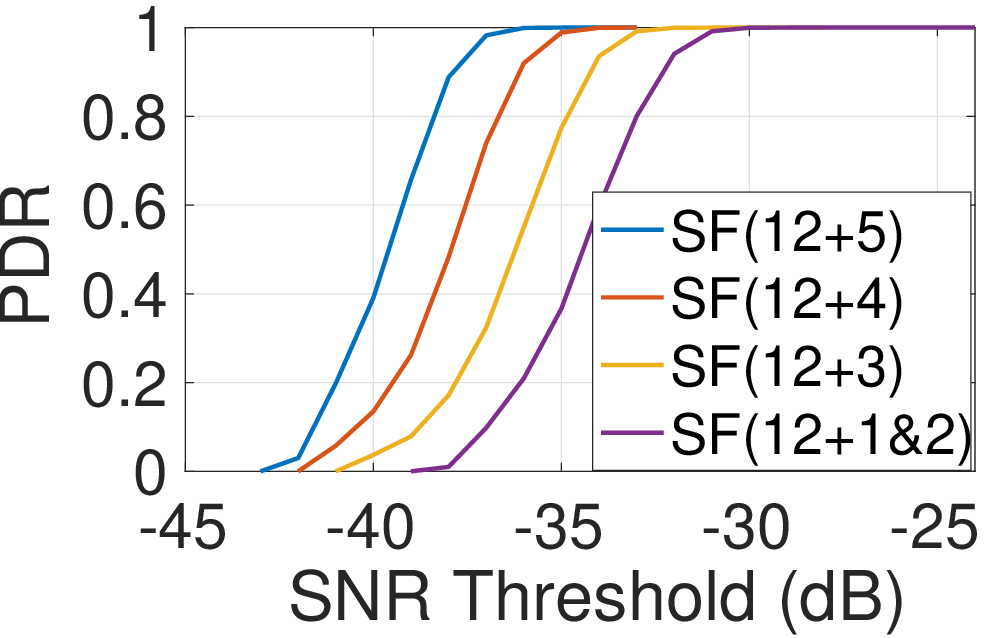}
		\label{fig:sf_prr_crs1}
	}
 \subfigure[]{
		\includegraphics[width=0.47\linewidth]{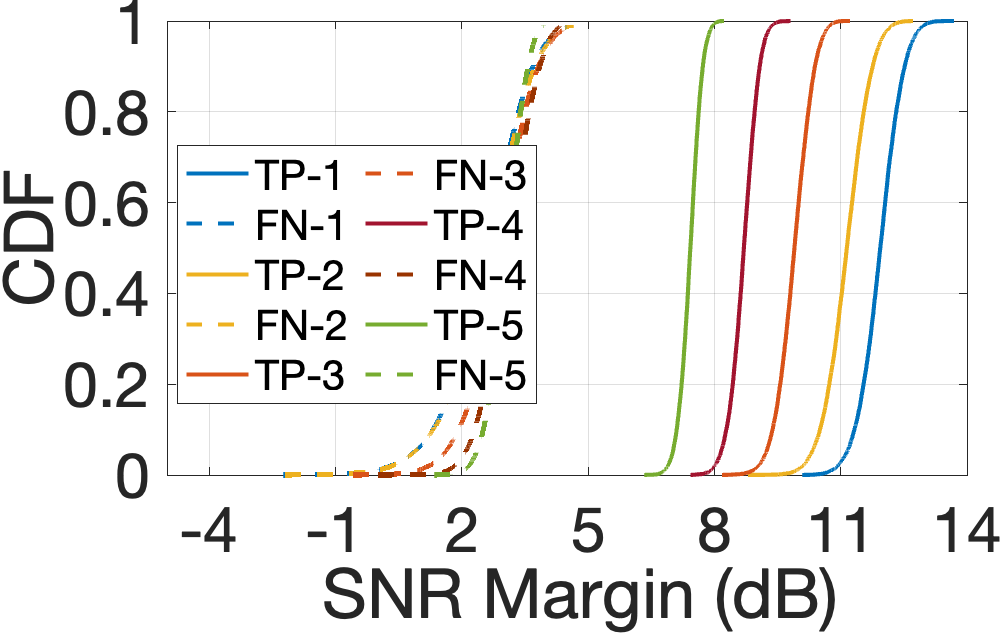}
		\label{fig:sf_prr_crs2}
	}
     
    \caption{PDR and SNR threshold illustration. (a) Packet Detection Rate (PDR) for different quasi-SF values. (a) SNR margin distribution for true positives and false negatives. TP-$m$ represents the CDF of SF$(12+m)$ packets when an actual packet arrives, while FN-$m$ represents the CDF of the Peak-to-Average-Ratio (PAPR) of the noise when no packet arrives. }
    \label{fig:prr_crs}
\end{figure}

Based on the SNR threshold, we plot the cumulative distribution function (CDF) of the SNR margin for both true positive detections and false negatives. A true positive is defined as the SNR margin for which two consecutive SF$(k+m)$ preamble chirps are detected. A false negative is defined as the minimum SNR margin at which the highest noise bin from two consecutive SFusion chirps exceeds the average noise level. This is also referred to as the Peak-to-Average Ratio (PAPR) of the noise. The SNR margin distributions for different quasi-SF values are shown in Fig.~\ref{fig:sf_prr_crs2}. TP-$m$ refers to the CDF of SF$(12+m)$ packets when an actual packet arrives, and FN-$m$ refers to the CDF of the PAPR of noise when no packet arrives.

From the figure, we see that the minimum SNR margin for true positive detection is typically two to three dB higher than that for false negative detection. This suggests that the proposed detection approach can successfully identify extremely weak SFusion packets amidst strong environmental noise. Moreover, since the SNR margin must lie between the minimum true positive and the maximum false negative values, we select the highest value that satisfies both conditions as the SNR margin for each quasi-SF. These values are listed in Table~\ref{tab:my_label}.

\begin{table}
    \centering
    \begin{tabular}{c|c|c|c|c|c}
        \hline
        & SF(12+1) & SF(12+2) & SF(12+3) & SF(12+4) & SF(12+5) \\
         \hline
      SNR  & -27 & -29 & -31 & -33 & -35\\  \hline
      TP & 10 & 8.5 & 8 & 7 & 6\\  \hline
      FN  & 6.5 & 6.5 & 6 & 5 & 4.5\\  \hline
      $\Gamma_1$    & 10 & 8.5 & 8 & 7 & 6\\  \hline
    \end{tabular}
    \caption{SNR Margin for Packet Detection}
    \label{tab:my_label}
\end{table}

\subsubsection{Adaptive PILOT insertion}
We next examine the optimal pilot insertion strategy under varying quasi-SF values and SNR levels. The pilot insertion rule is determined by the minimum number of pilots required to maintain a bit error rate (BER) below 
$10^{-4}$ without applying the opportunistic decoding strategy.

\begin{figure}
\centering
\includegraphics[width=0.95\linewidth]{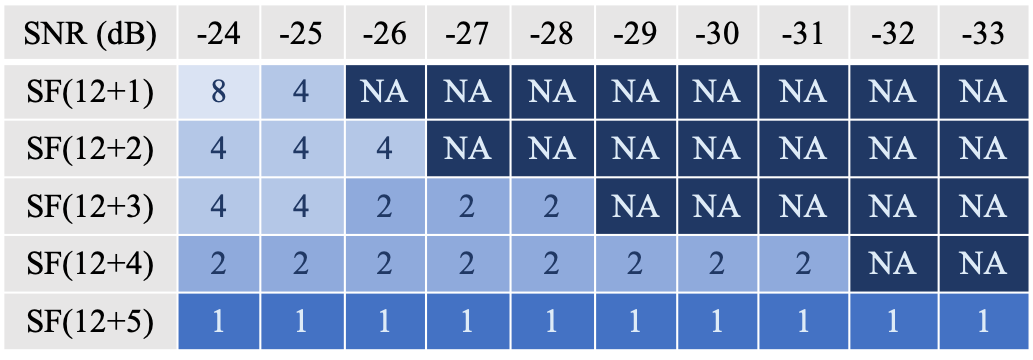}
\vspace{-0.1in}
    \caption{The optimal pilot interval under different SNR.}
\label{fig:Pilot_insert}
\vspace{-0.05in}
\end{figure}

As shown in Fig.~\ref{fig:Pilot_insert}, the optimal pilot interval varies with SNR. In general, higher SNR conditions require fewer pilot symbols, allowing for longer pilot intervals. For example, when transmitting quasi-SF$(12+1)$ packets under SNR conditions at or below -24dB, the optimal pilot interval is 8, meaning one pilot is inserted every eight consecutive super symbol blocks.

It is also worth noting that within the SNR regime where both Ostinato and SFusion are operational—specifically, at SNR levels below -24dB according to Fig.~\ref{fig:baselinesBER}-Ostinato incurs a pilot overhead of 1/6, as it is restricted to LDRO mode. In contrast, SFusion requires only a 1/32 overhead in this regime. This highlights that, by carefully compensating for hardware offsets, SFusion not only significantly extends the operational SNR range (as we will detail later), but also dramatically reduces pilot overhead and improves transmission efficiency.



\subsubsection{SNR performance with grouped repetition code} \label{SNRperformance}
\begin{figure}
\centering
\includegraphics[width=0.85\linewidth]{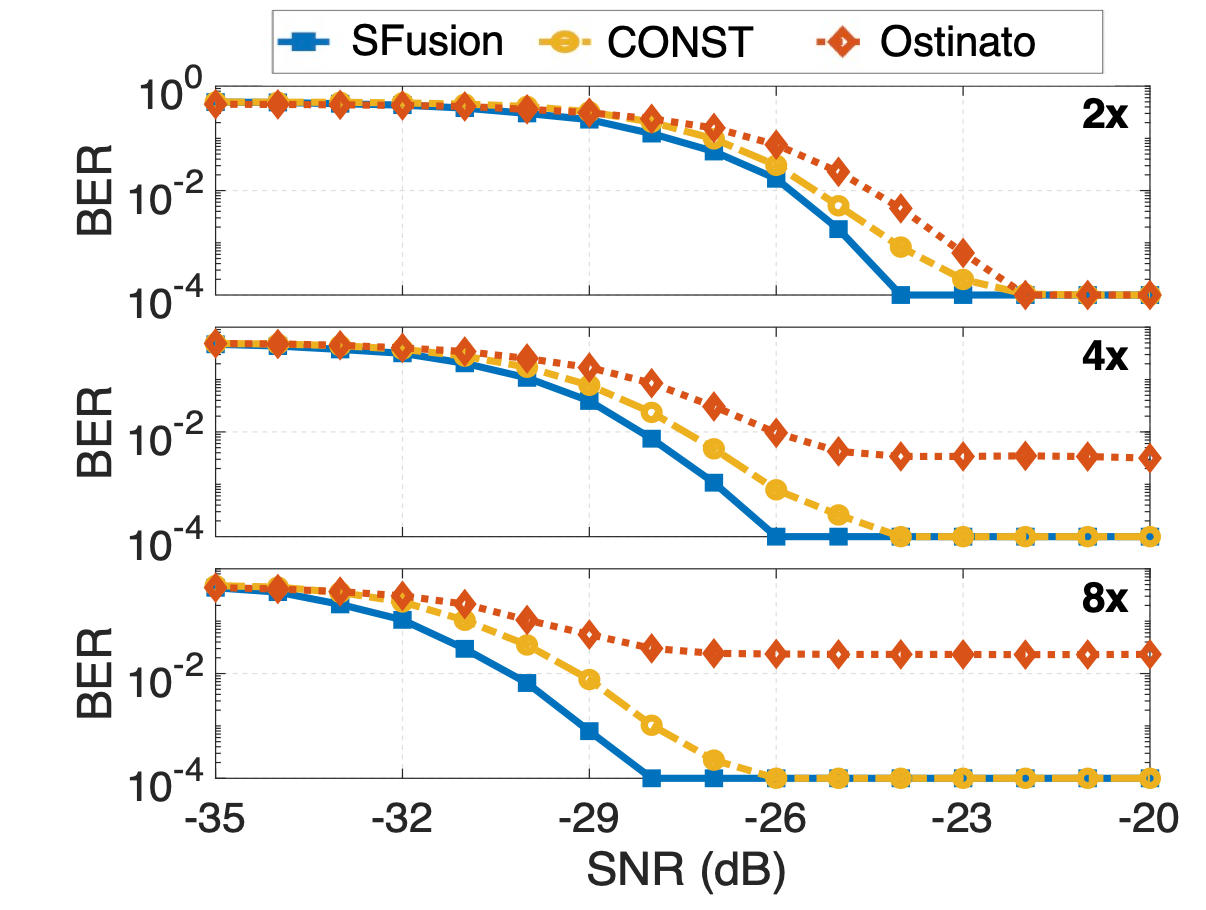} 
\vspace{-0.1in}    
\caption{Bit Error Rate (BER) versus SNR performance of SFusion, Ostinato and CONST with 2x, 4x, and 8x repetitions respectively. LDRO mode is enabled in Ostinato.}
\label{fig:baselinesBER}
\vspace{-0.15in}
\end{figure}

Once pilots are well arranged, we then study the SNR performance without the proposed opportunistic decoding scheme. Fig. \ref{fig:baselinesBER} compares the BER versus SNR performance of SFusion to that of Ostinato and CONST with various repetition rates (for CONST, gateways). Here, to study Ostinato's repetition performance, repetition factor $K=8$ is implemented by setting all payload symbols to zero. However, this mode cannot be used to transmit meaningful messages. 

According to the results, given double repetitions, SFusion achieves a 2dB SNR gain compared with Ostinato and CONST. Moreover, when repetition rate increases, the combining performance of SFusion and CONST scales smoothly as signal offsets are properly compensated for. SFusion achieves a better combining performance of 2dB compared with CONST, because the PILOT strategy used in SFusion estimates the signal offsets by known symbols. CONST, on the other hand, leverages error-prone random symbols for offset estimation, leading to less accurate offset compensation. Ostinato suffers from constant bit errors when its repetition factor $K \geq $ 4. This is because a high repetition rate generates long packets, leading to a frequency shift greater than $16 / 3$ bins at the packet level, which exceeds the tolerable range of LDRO mode.

\begin{figure}
\centering
\includegraphics[width=0.8\linewidth]{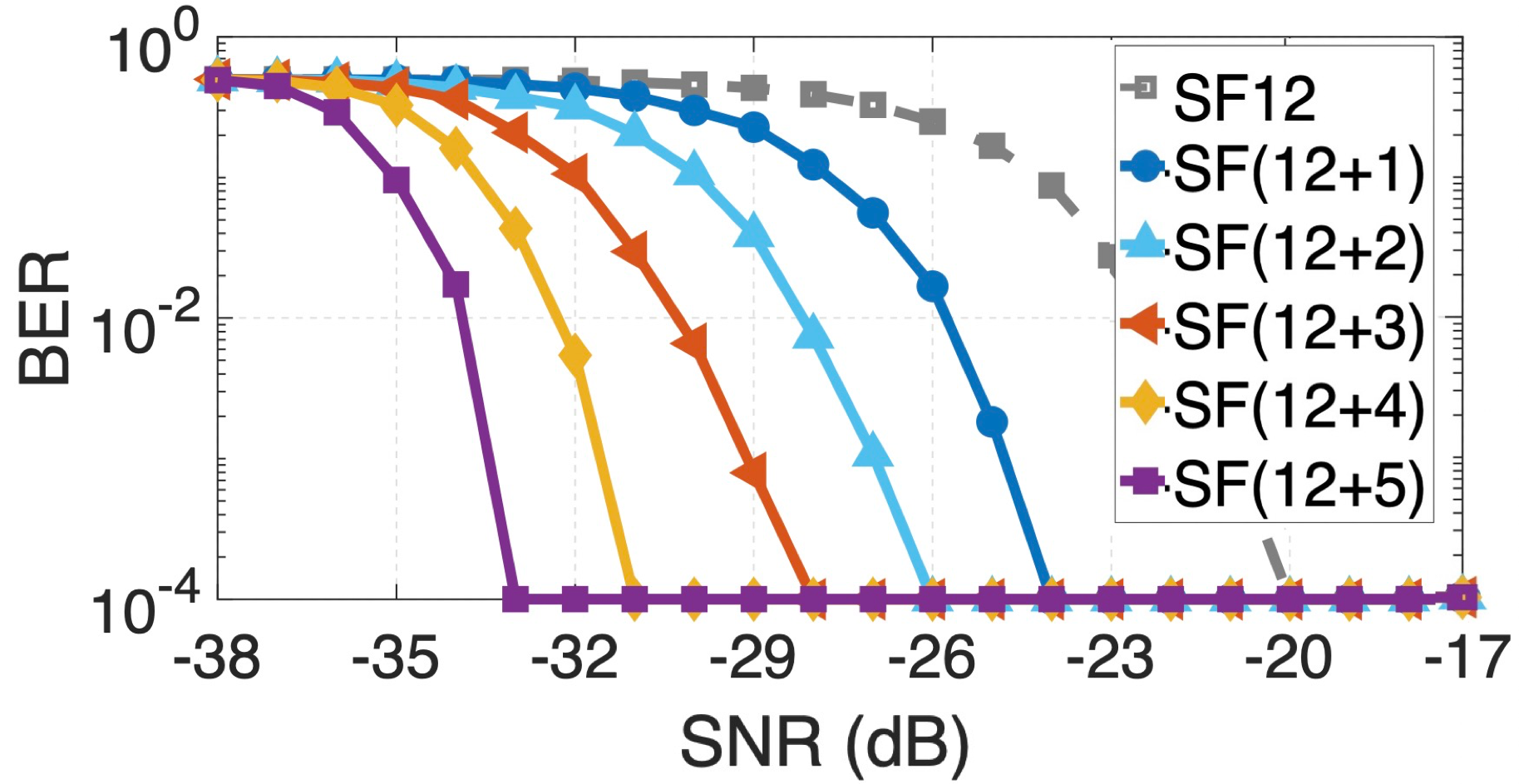}
\vspace{-0.05in}
    \caption{BER versus SNR performance of SFusion.}
\label{fig:SFanyBER}
\vspace{-0.1in}
\end{figure}


Fig. \ref{fig:SFanyBER} further illustrates the performance of SFusion with different quasi-SF values. Given BER $\leq 10^{-4}$, the supportable SNR for quasi-SF$(12+5)$ packet without opportunistic decoding approaches is -33dB, an SNR improvement of 13dB compared with traditional SF12 packet, and an SNR improvement of more than 11 dB compared with Ostinato. This improvement in combining performance is due to the meticulous compensation for both intra-symbol phase offsets and inter-symbol frequency drifts. As a result, SFusion can tolerate different packet-level frequency drifts, leading to a seamless SNR improvement with increasing repetition rates. Therefore, we argue that SFusion can be extended to large repetition rates, thus effectively enhancing the SNR of an extremely weak link.


\begin{figure}
\centering
    \includegraphics[width=0.85\linewidth]{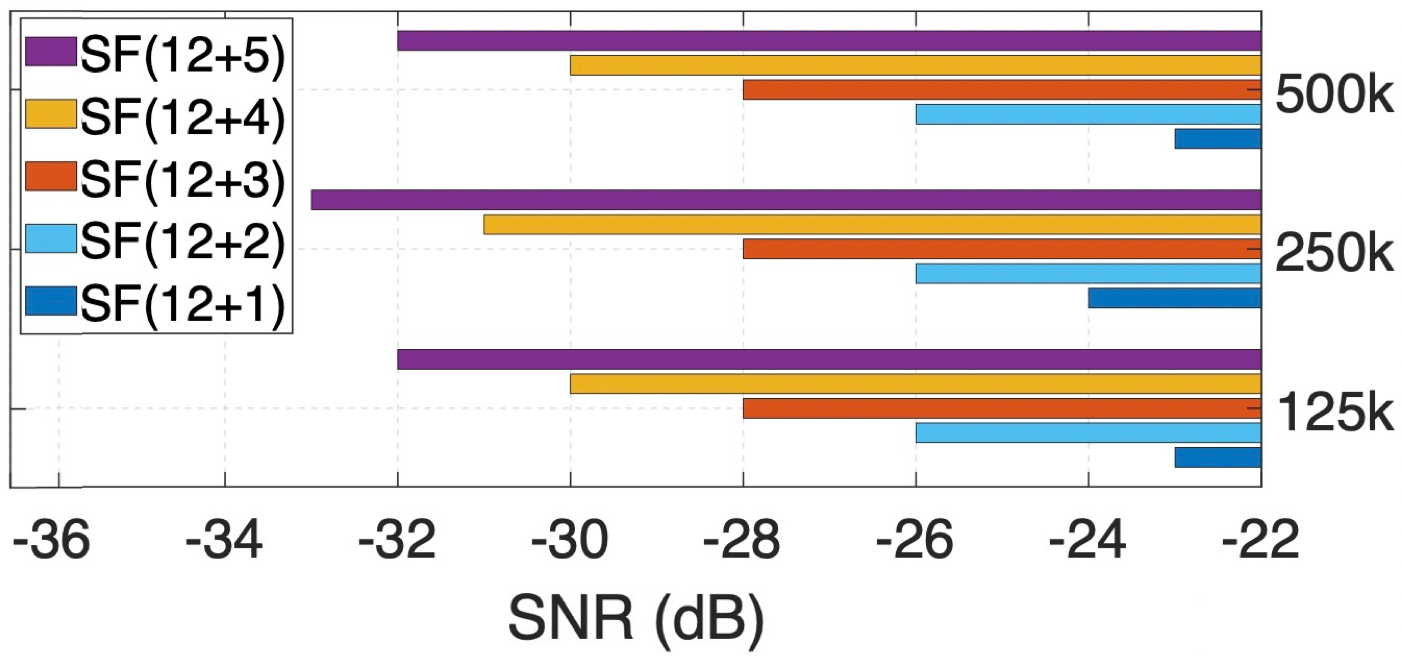}
   
\vspace{-0.05in}
\caption{The lowest achievable SNR with the grouped repetition code under different bandwidths given BER $\leq 10^{-4}$.}
\label{fig:SFanyBW}
\end{figure}

\begin{figure}
    \centering
    \includegraphics[width=0.95\linewidth]{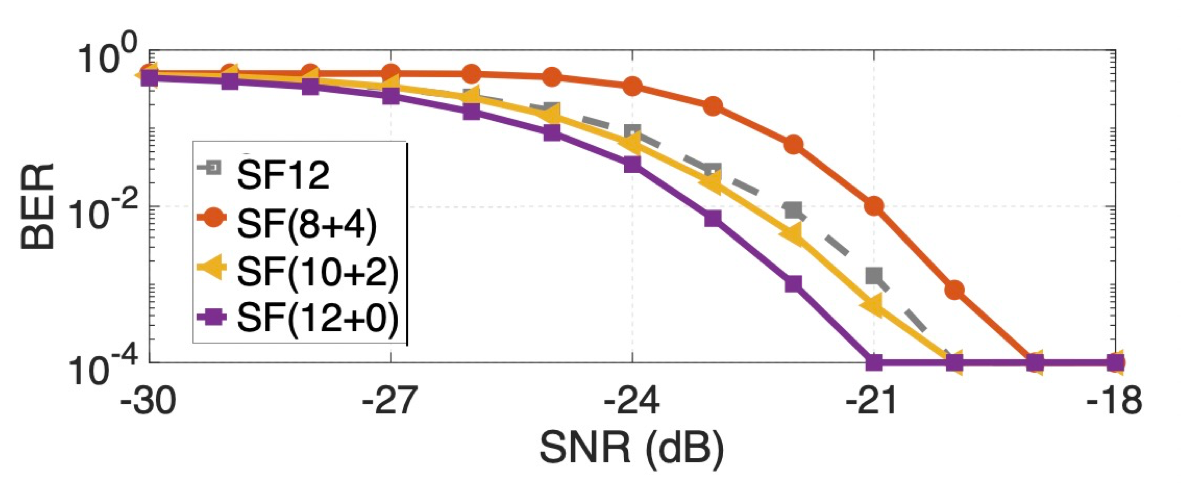}
\vspace{-0.1in}
    \caption{BER versus SNR performance of standard SF12 symbol and simulated quasi-SF12 symbol using $2^0$ SF12 symbols, $2^2$ SF10 symbols and $2^4$ SF8 symbol respectively.}
    \label{fig:sf12simulation}
    \vspace{-0.1in}
\end{figure}
We further examine the impact of bandwidth on SFusion's performance. Fig. \ref{fig:SFanyBW} presents a robust scaling performance with various quasi-SF values under different bandwidths, with an up to 1dB performance deviation for the same repetition rate. For each bandwidth setting, given BER $\leq 10^{-4}$, we observe a 2dB to 3dB SNR improvement as the repetition rate doubles. This result demonstrates that SFusion can be effectively adopted for various bandwidth.

Finally, to verify SFusion's capability of versatile transmission rate, we simulate an SF12 symbol with $2^0$ SF12 symbol, $2^2$ SF10 symbols, and $2^4$ SF8 symbols (denoted by quasi-SF$(8+4)$, quasi-SF$(10+2)$, quasi-SF$(12+0)$ respectively). We then compare the SNR performance of SFusion symbols with that of standard SF12 symbols. As illustrated in Fig. \ref{fig:sf12simulation}, given BER $\leq 10^{-4}$, the supportable SNR limit of standard SF12 symbols is -20dB. quasi-SF symbols achieve a combining performance from -19dB to -21dB, comparable to that of SF12 symbol in LoRa PHY. Moreover, the quasi-SF$(12+0)$ outperforms SF12 of LoRa PHY by 1 dB by compensating for signal offsets. The SNR difference between each pair of quasi-SF packets is caused by different number of coherently combined chips inside each symbol. E.g, the SF$(8+4)$ scheme coherently combines 256 chips each time and directly adds up the results, while SF$(12+0)$ coherently combines all the 4096 chips, leading to a 2dB gain. This result demonstrates that SFusion can effectively simulate high-SF symbols by combining symbols of lower SF values, thus supporting versatile transmission rate.

\subsubsection{SNR performance with opportunistic decoding}
\begin{figure}
\centering
    \includegraphics[width=0.95\linewidth]{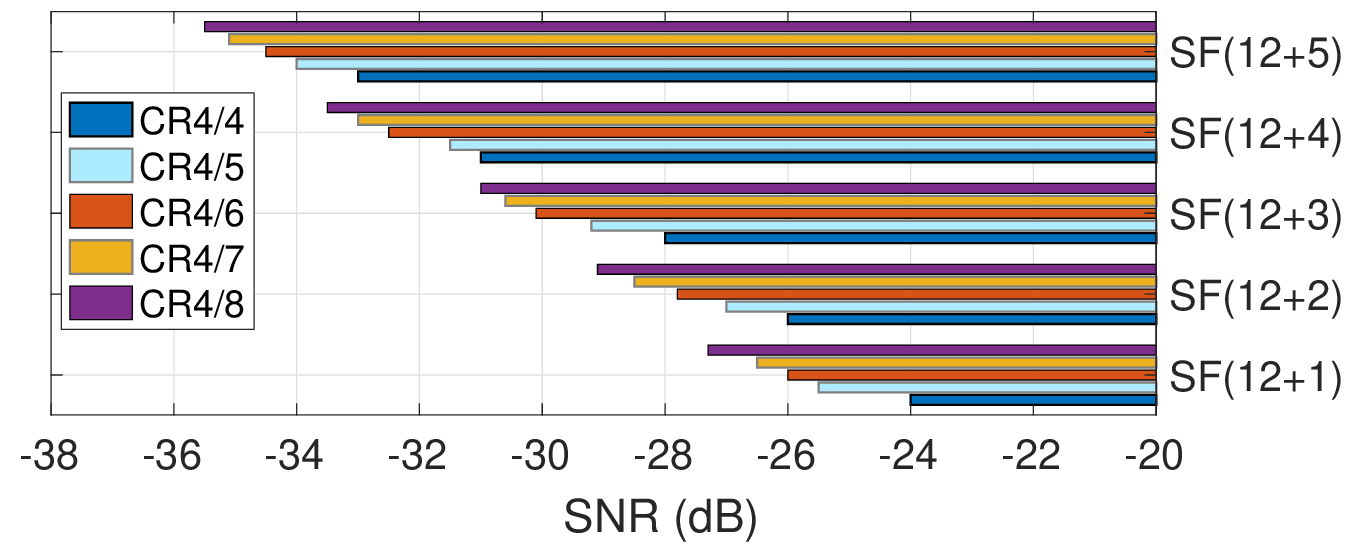}
\vspace{-0.1in}
    \caption{SNR improvement brought by the joint demodulation and decoding scheme given BER $\leq 10^{-4}$.} 
    \label{fig:SNRimprovement}
\vspace{-0.05in}
\end{figure}

Based on the grouped repetition code, the proposed opportunistic decoding scheme further leverages the coding correlations across symbols in a block to extend SFsuion's SNR limit. Fig. \ref{fig:SNRimprovement} shows the supportable SNR for various quasi-SF values and various coding rates given BER $\leq 10^{-4}$. According to the figure, the higher the coding rate, the higher the redundant information to be exploited for the decoding scheme, thus the higher the SNR improvement. Overall, the proposed decoding scheme achieves an improvement of 1dB to 3dB SNR gain compared to the case when no decoding scheme is adopted.

\begin{figure}
    \centering
    \includegraphics[width=0.95\linewidth]{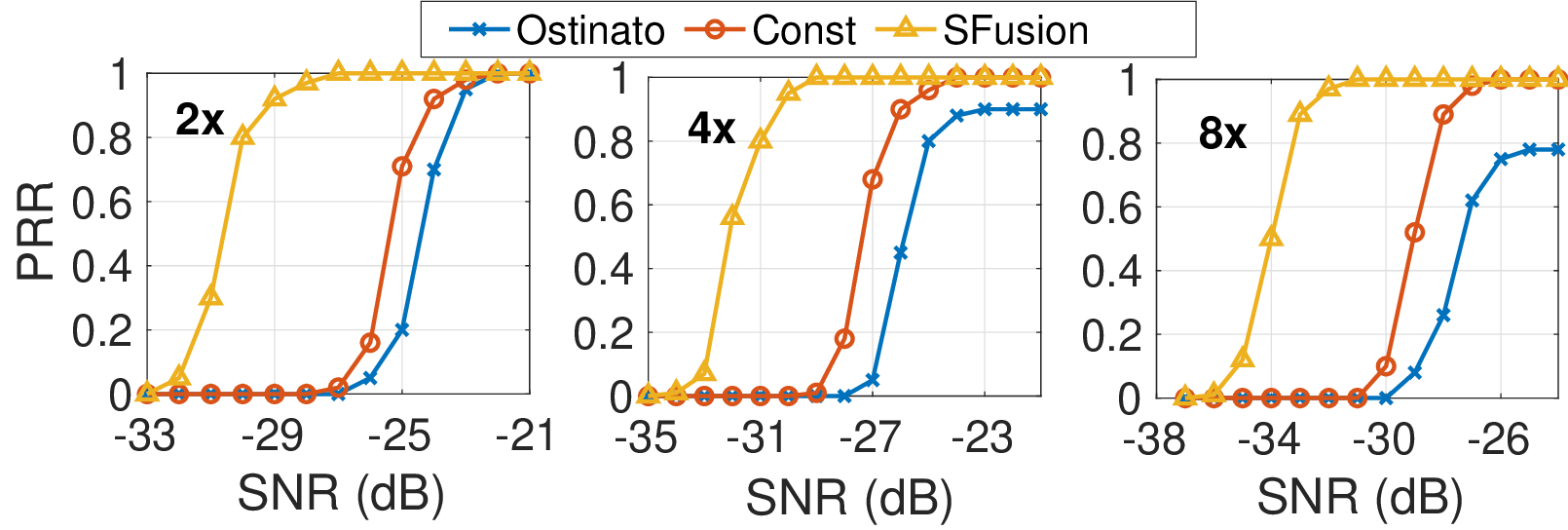}  
    \vspace{-0.1in}
    \caption{Packet Reception Rate (PRR) versus SNR performance of SFusion (CR4/8), Ostinato and CONST with 2x, 4x, and 8x repetitions respectively.}
    \label{fig:PRRbaselines}
    \vspace{-0.2in}
\end{figure}

Fig. \ref{fig:PRRbaselines} compares the overall PRR (Packet Reception Rate) of Ostinato, CONST and SFusion with different repetition rates. Here, SFusion adopts CR4/8 coding. SFusion outperforms CONST and Ostinato by 6dB to 7dB with the same repetition rate. Ostinato, though remains effective for a lower repetition rate (i.e., $K =$ 2) as it yields a 2dB improvement on PRR compared with LoRa, becomes inefficient with higher repetition rates. This inefficiency arises as higher repetition rates significantly increase the packet length, leading to greater packet-level frequency drift.

\begin{figure}
    \centering
    \includegraphics[width=0.95\linewidth]{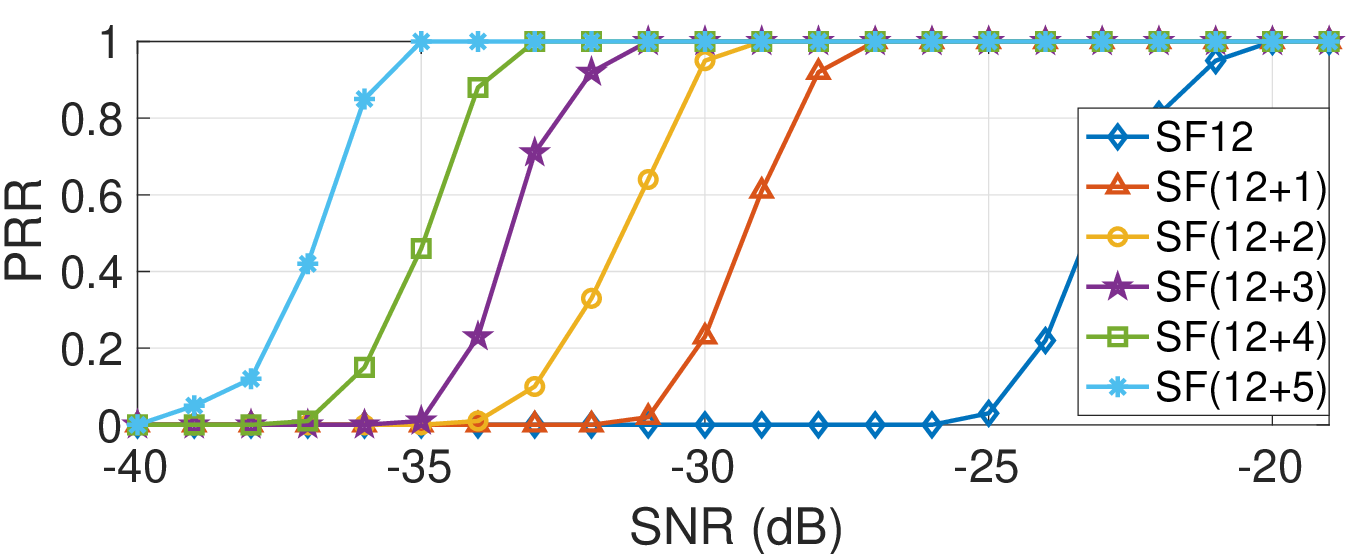}
    \vspace{-0.1in}
    \caption{PRR versus SNR performance of SFusion (CR4/8) and SF12 of LoRa PHY.}
    \label{fig:SFanyPRR}
\vspace{-0.1in}
\end{figure}

Fig. \ref{fig:SFanyPRR} further depicts the overall PRR of SFusion as well as the standard LoRa SF12 packet. As the quasi-SF value increases, SFusion's SNR improvement scales smoothly by 2dB to 3dB. When SNR$=$-35dB, the packet reception rate with quasi-SF$(12+5)$ approaches 1, outperforms the LoRa SF12 packet by 15dB, which is a composed result of aggregated energy with meticulous compensation for signal offsets (14dB) and coding gain (2dB). 


\subsubsection{Combining SFusion with Multi-gateway Approach} 
\begin{figure}
\centering
\includegraphics[width=0.95\linewidth]{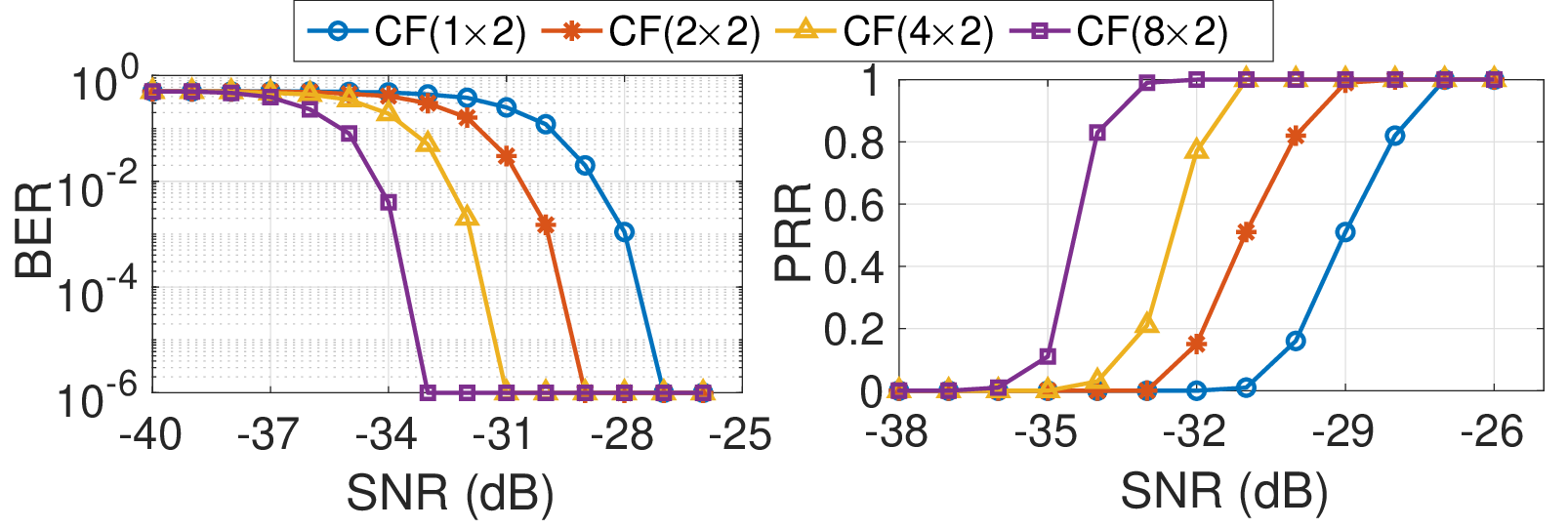}
\vspace{-0.1in}
\caption{BER and PRR Performance of CONST-Fusion.}
\label{fig:constany}
    \vspace{-0.1in}
\end{figure}

We evaluate SFusion's performance when combining with the orthogonal method CONST. The combination approach is named \textbf{CONST-Fusion}, and abbreviated as CF for short. As depicted in Fig. \ref{fig:constany}, CF is formatted as the number of gateways $\times$ the repetition rate. For instance, CF$4\times 2$ refers to the scenario that four gateway nodes jointly receive quasi-SF(12+1) packets. 

Given double repetitions, the combined performance scales smoothly with the number of gateways, achieving 2dB gain each time the number of gateways doubles. Even under the SNR of -33dB, CONST-Fusion$(2\times 8)$ achieves BER of $\leq 10^{-4}$, outperforming CONST-Fusion$(2\times 1)$ by 6dB. These results indicate SFusion can be combined with orthogonal methods to perform better.

\subsubsection{Outdoor experiments} 
\begin{figure}
\centering
    \includegraphics[width=0.9\linewidth]{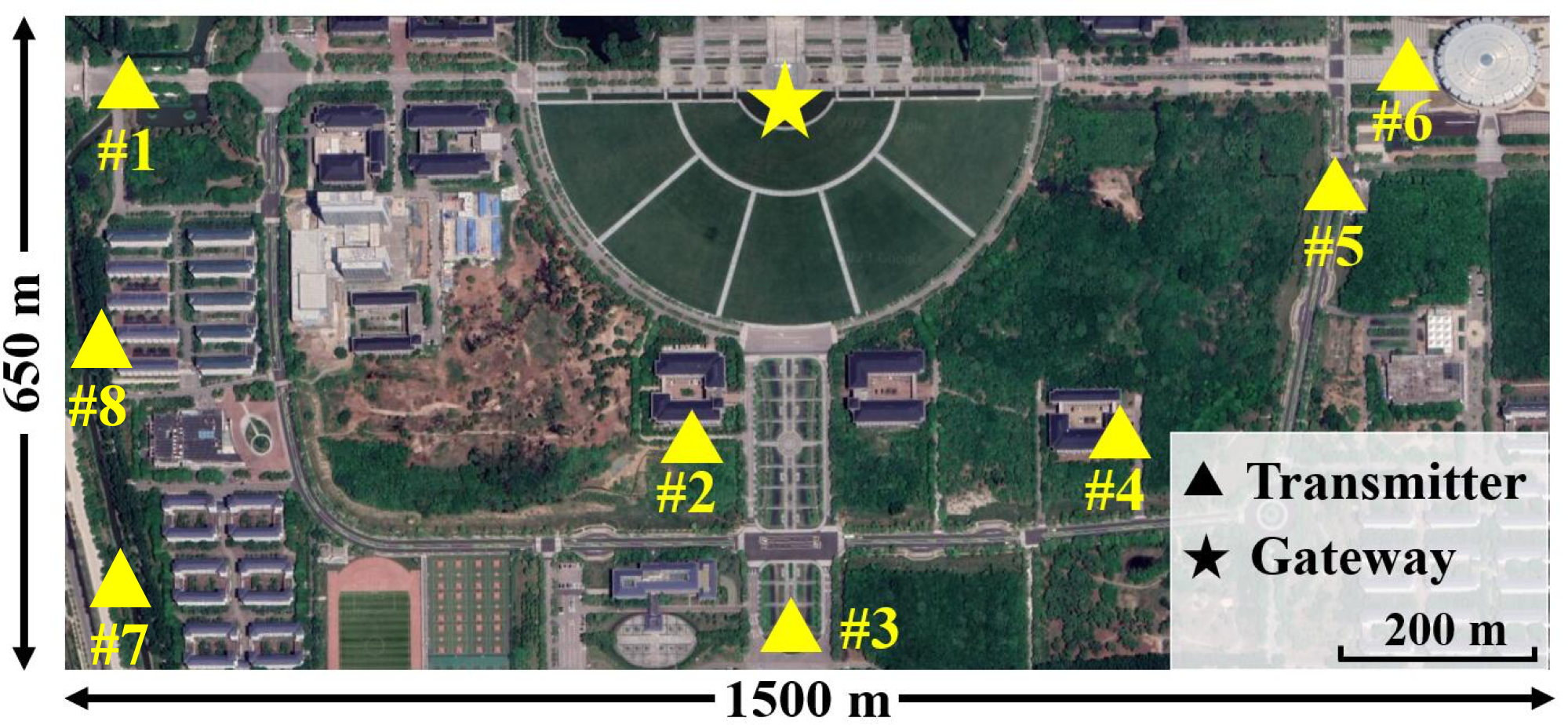}
    \vspace{-0.1in}
    \caption{Deployment of the gateway and transmitters.}
    \label{fig:outdoorenv}
    \vspace{-0.15in}
\end{figure}

We next perform outdoor experiments in a 1500 meters $\times$ 650 meters campus area (shown in Fig. \ref{fig:outdoorenv}). The gateway is denoted with a yellow star in the figure, and the transmitting nodes (labeled with yellow triangles) are scattered around the gateway. The maximum communication range is about 1000 meters. While some nodes are situated in regions with a clear line of sight path to the gateway, the connections of other nodes to the gateway are hindered by buildings, trees, and various impediments. As a result, different links, even when at a similar distance from the gateway, encounter distinctly varied channel conditions.

Specifically, we choose 8 positions for this outdoor experiment, among which position \#1, \#3 and \#6 can establish LOS links with the gateway, position \#2 and \#7 and \#8 are blocked by buildings, position \#4 is inside one building, and position \#5 is blocked by a long stretch of woods. We keep the basic experimental parameters same as in indoor experiment, and compare the PRR and throughput performance of SFusion with Ostinato and the standard LoRa packet (SF12). We use quasi-SF$(12+1)$ with CR4/5. The repetition rate $K$ of Ostinato is set to 2 and the LDRO mode is enabled. 

\begin{figure}
\centering
\includegraphics[width=0.95\linewidth]{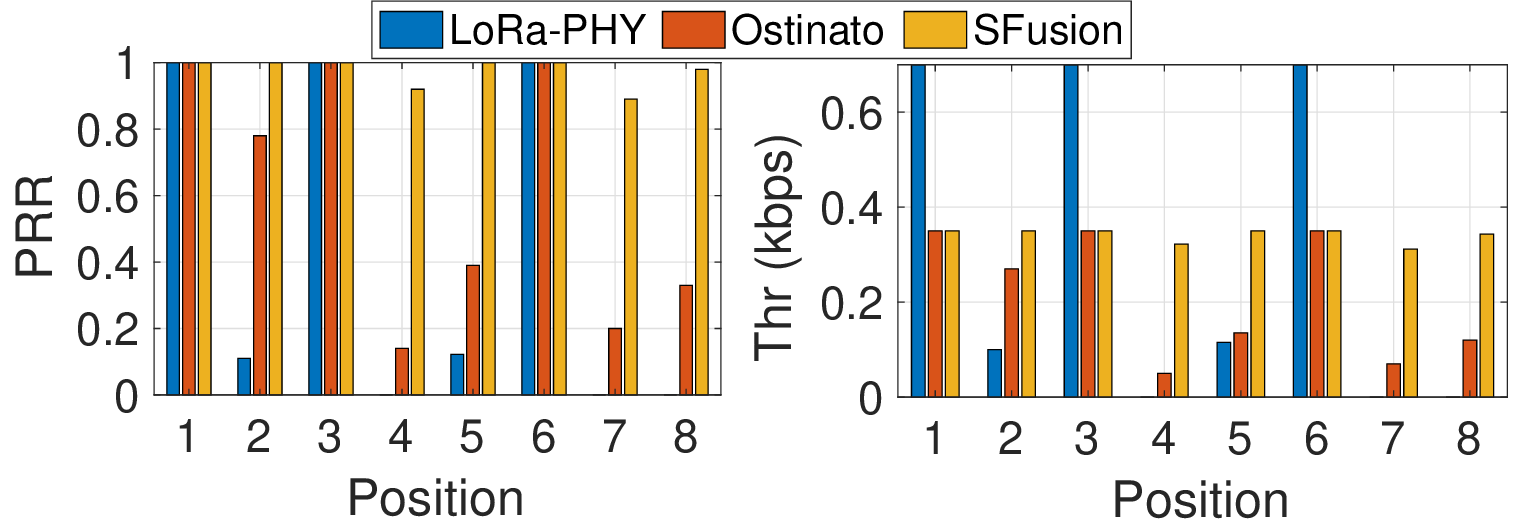}
    \vspace{-0.1in}
    \caption{Performance of SFusion in outdoor environments. (a). PRR performance and (b). throughput performance at different positions.}
    \label{fig:realenv}	
    \par
    \vspace{-0.15in}
 \end{figure}

As shown in Fig. \ref{fig:realenv}(a) in position \#1, \#3 and \#6, all three schemes achieves perfect PRR performance since LoRa inherently takes a great advantage at long range, especially for LOS communication. In each place, the SNR fluctuates hugely from -15dB to -4dB, partly caused by moving objects such as trucks, cars and passengers. 

When obstructed by two buildings at position \#2, standard LoRa SF12 packet achieves a PRR of 12\% at an SNR of approximately -24dB. Ostinato experiences a 20\% PRR degradation (PRR=80\%), while SFusion maintains a PRR of 100\%. At position \#4, the SNR decreases to around -28dB, as multiple concrete walls block the LoRa communication. Due to the weak SF12 symbol energy, LoRa's functionality is severely compromised, with PRR dropping to zero. Ostinato successfully receives 13\% of the packets, while SFusion achieves around 92\% PRR. The primary reason for the decline in PRR for Ostinato is its ineffective signal combining approach. At position \#5, While woods obstruct the line of sight, they do so to a lesser extent than concrete walls, leading to a similarly low SNR (around -24dB) but higher channel variations compared to position \#2. PRR performance for standard LoRa and Ostinato is 12\% and 37\%, respectively. SFusion consistently achieves over 90\% PRR for all outdoor transmissions. Ostinato, while improving SNR for some links, exhibits significantly lower PRR for links with SNR $\leq$ -24dB. At position \#7 and \#8, the due to the long distance and obstructed buildings, the SNR also approaches to -28dB, leading to severe packet reception failure when traditional LoRa-PHY is adopted to received the packet. Ostinato, though is able to receive 20\% to 33\% of the packet, still suffers from significant packet loss. However, SFusion keeps showing over 90\% packet reception rate in these cases.

Fig. \ref{fig:realenv}(b) shows the effective throughput of the three schemes. Here effective throughput refers to the number of successfully received payload bits per second. Traditional LoRa-PHY undoubtedly achieves the highest throughput in LOS scenarios, as it does not suffer from PRR loss and has no repetition cost. Osinato and SFusion achieve about half the throughput as LoRa. Nevertheless, when SNR is low, the advantage of LoRa-PHY on transmission rate vanishes as it suffers from a poor PRR. The throughput of Ostinato and SFusion is better because they are more resilient to noise, while SFusion is notably more robust and remains stable consistently. 

%% file: sections/related_work.tex
\section{Related Works}\label{relatedworks}
Communication on weak links is an important issue in reliable LoRa communications. Generally, current works can be classified into two categories: (a), enhancing signal quality in physical layer, and (b), repairing corrupted information from coding redundancy.

\textbf{Methods for enhancing signal quality.} These methods exploit the diversity of received signals, constructively adding up these signals helps to average out the noise while enhancing target signals. Charm \cite{dongare2018charm} collects multiple signal copies from different gateways and combines them in the cloud. However, it suffers from high edge-cloud communication costs and non-coherent signal combinations. Nephalai \cite{liu2020nephalai} designs a compressive sensing algorithm to compress IQ signals, alleviating upload pressure. CONST \cite{zhang2022const} coherently combines signals from different gateways, suppressing the power of noise rapidly while aggregating the signal power. MALoRa \cite{hou2023don} recovers weak signals by coherently combining signals received by synchronized antennas. Choir \cite{eletreby2017empowering} suggests simultaneously sending similar data by multiple transmitters to improve the received energy. However, synchronization among transmitters is a great challenge. Ostinato \cite{xu2022ostinato} and Xcopy \cite{xia2023xcopy}, however, leverage the time diversity for signal enhancement, avoiding additional hardware costs. However, as discussed, Ostinato struggles to scale due to stringent hardware imperfections. Xcopy resolves low-SNR packets reception problem via multiple retransmissions. This incurs extremely long delays under the 1\% duty cycle imposed by EU regulation \cite{rp002}, potential collisions due to the cumulative length of all retransmitted packets, and violation of LoRa MAC-layer protocol. In contrast, SFusion implements repetition coding within one packet, which supports the repetition rate up to 32 given the max length of 255 Bytes specified in LoRa PHY \cite{semtech2020sx1276}. This not only avoids the long delay and extra overhead of retransmissions but also complies with LoRa MAC-layer protocol. Moreover, SFusion harnesses the time diversity, provided by the repetition within a long packet, for better demodulation and decoding.

\textbf{Methods based on coding.} These methods recover corrupted information from coding redundancy. DaRe \cite{marcelis2017dare} designs a combination of Convolution code and Fountain code to perform rateless coding. OPR \cite{balanuta2020cloud} votes for payload bits with RSSI information for packet received by multiple gateways. Similar to OPR, ECRLoRa \cite{mei2023ecrlora} selects candidate packet segments at gateway, performing group weighted voting to recover corrupted data. LLDPC \cite{yang2022lldpc} designs a low density parity check coding scheme to build constraints among payload bits, and decodes with the LLR information using a graph neural network. DC \cite{yazdani2022divide} combines Reed-Solomon codes with LoRa FEC mechanism to protect error-prone bytes in a LoRa packet. These coding methods struggle at extremely low SNR levels as noise corrupts most symbols, leaving little usable redundancy.

Beside obtaining various gains from extra hardware or protocol designs, NELoRa \cite{li2021nelora} designs a Deep Neural Network to distinguish the time-frequency pattern of LoRa symbol from background noise in spectrumgram. Falcon \cite{tong2021combating} deliberately interferes or maintains on-going LoRa packets with another transmitter to modulate binary bits, completely sacrificing data rates. A similar work \cite{tong2022spreading} also leverages coherent interference among signals to support long-range cross-technology communication.

%% file: sections/conclusion.tex
\section{Conclusion}\label{conclusion}
In this paper, we present SFusion, a software-based coding paradigm, that enables modulating a LoRa packet with any spreading factor and compatible with orthogonal methods, facilitating reliable city-wide communications. SFusion employs $2^m$ SF$k$ symbols to emulate a quasi-SF$(k+m)$ symbol, incorporating pilot symbols to compensate for time-varying frequency drifts and phase offsets. It also introduces a grouped repetition code for SFusion symbol modulation and utilizes a joint demodulation and decoding scheme to seamlessly integrate energy distribution into the decoding process. This transformation effectively strengthens the Hamming decoding, significantly improving packet reception rates, especially for ultra-low SNR packets. Extensive evaluations showcase SFusion's exceptional improvement by up to 14dB over SF12 of LoRa PHY. Moreover, SFusion outperforms the leading solutions for PHY layer reliable communications, achieving an improvement of up to 13dB compared with Ostinato, and 6dB to 7dB improvement over CONST with same repetitions. Importantly, SFusion is the first work to consistently deliver high performance across various spreading factors.